\documentclass[10pt]{iopart}

\usepackage{graphicx}
\begin{document}

\review[Physics with coherent matter waves]{Physics with coherent
matter waves}

\author{Kai Bongs and Klaus Sengstock
\footnote[3]{To whom correspondence should be addressed
(sengstock@physnet.uni-hamburg.de)} }

\address{Institut f\"ur Laser-Physik, Universit\"at Hamburg,
Luruper Chaussee 149, 22761 Hamburg, Germany}

\begin{abstract}
This review discusses progress in the new field of coherent matter
waves, in particular with respect to Bose-Einstein condensates. We
give a short introduction to Bose-Einstein condensation and the
theoretical description of the condensate wavefunction. We
concentrate on the coherence properties of this new type of matter
wave as a basis for fundamental physics and applications. The main
part of this review treats various measurements and concepts in
the physics with coherent matter waves. In particular we present
phase manipulation methods, atom lasers, nonlinear atom optics,
optical elements, interferometry and physics in optical lattices.
We give an overview of the state of the art in the respective
fields and discuss achievements and challenges for the future.
\end{abstract}



\maketitle

\section{Introduction}
Quantum mechanics as an important foundation of modern physics and
its description in terms of probability densities naturally
incorporates the fascinating wave nature of massive particles.
These matter waves were postulated in 1925 by de
Broglie~\cite{deBroglie1925} and demonstrated for electrons by
Davisson and Germer~\cite{Davisson1927}. Matter waves and their
ability to interfere have been and still are the basis for many
fundamental tests of the principles of quantum mechanics. In
applications the interference of atomic matter waves nowadays sets
the basis of time and allows the construction of some of the most
sensitive sensors, e.g. for gravity, rotation, gravity gradients,
atom polarisabilities, ...
\cite{Borde1989a,Carnal1991a,Keith1991a,Riehle1991a,Kasevich1991a,Weiss1993a,Gustavson1997a,Berman1997a,Snadden1998a,Gustavson2000a}.

The realisation of Bose-Einstein condensation in dilute atomic
gases in 1995~\cite{Anderson1995a,Davis1995b,Bradley1995a} opened
an exciting window to quantum mechanics and started a new era of
matter wave research. The impact of this disovery was underlined
by the Nobel price 2001 awarded to the pioneering
groups~\cite{Cornell2002a,Ketterle2002a}. Until this time all
experiments were based on single particle interference with
thermal ensembles of particles having a coherence length typically
much below a $\mu $m. The phenomenom of Bose-Einstein condensation
allows for the first time the creation of a macroscopically
occupied matter wave with coherence length up to the mm scale. A
Bose-Einstein condensate wavefunction compares to a thermal
ensemble in an analog way as a laser in optics to a light bulb.
The coherence and the macroscopic nature of this new type of
matter wave have paved the way for a new field of physics and many
fascinating experiments. As an example in analogy to laser
physics, the interference of two individual Bose-Einstein
condensate wavefunctions~\cite{Andrews1997b} clearly demonstrated
multi-particle interference with matter waves, which could not be
observed with thermal ensembles. In this article we will
concentrate on the coherence aspects of the Bose-Einstein
condensate wavefunction and discuss a selection of relevant
experiments in the field.

The history of Bose-Einstein condensation goes back to the
''golden time'' of quantum mechanics, when in 1924 Bose developed
a new, statistical approach to derive Planck's law for the
radiation of a black body~\cite{Bose1924a}. Due to publication
problems Bose asked Einstein for advice, who recognised the
importance of this work, supported the publication and already
then added a note, that this new statistics could be extended from
photons to massive particles. Einstein published the resulting
statistics, later called ''Bose-Einstein statistics'' in
1925~\cite{Einstein1925a}, finding the solely statistically
motivated condensation of particles into the ground state of the
system without the involvement of interactions as an immediate
consequence. It is interesting to note that Einsteins original
proposal for the experimental realisation of this phase transition
was based on electrons as at this time the distinction between
''Bosens'' and ''Fermions'' was not yet made. Einsteins prediction
lead to many controverse discussions and at first was rejected by
most physicists, in particular as there seemed to be no
opportunity to observe this phenomenon in experiment.

For example, the BEC phase transition is usually masked by
interparticle interactions, which for atomic gases lead to a gas
to fluid or gas to solid transition, as it lies in a
thermodynamically forbidden region of the phase diagram. The first
experimental evidence of Bose-Einstein condensation was found by
F. London in 1938~\cite{London1938a,London1938b}, in analysing the
superfluid state of liquid Helium. This connection was
controversely discussed for a long time and hard to verify in
experiment due to the interactions in He, which limit the
Bose-Einstein condensate fraction to about 10\%. Experimental
attempts to observe Bose-Einstein condensation in a nearly ideal
gas first concentrated on spinpolarized
Hydrogen~\cite{Stwalley1976a,Silvera1980b,Hardy1982a,Hess1983a,Johnson1984a}
and later also on alkali atoms. It took 70 years from Einsteins
prediction until the combination of laser and evaporative cooling
techniques finally lead to success in weakly interacting gases in
1995. The use of a dilute gas at extremely low temperatures was
the key to these experiments as it allowed to work in a metastable
regime from the point of view of classical phase transitions
(needing 3-body collisions to form clusters) while having a
thermalized gas phase (relying on two-body collisions).

The physics of BEC was already summarized in several review
articles,
e.g.~\cite{Cornell1996a,Burnett1999a,Cornell1999a,Ketterle1999a,Dalfovo1999a,Griffin1999b,Helmerson1999b,Courteille2001a,Stamper-Kurn2001a,Ketterle2001a,Cornell2002a,Cornell2002b,Bongs2003a}
and textbooks,
e.g.~\cite{Griffin1995c,Martellucci2000a,Savage2001a,Pethick2002a,Pitaevskii2003a}.

The intention of this article is to give an introduction to the
coherent matter wave aspects intrinsically related to
Bose-Einstein condensates. The remainder of this section will give
an overview of the theory of Bose-Einstein condensation and
condensates in ideal and in dilute gas systems.

\subsection{Theoretical description of BEC}

\subsubsection{Statistics of the BEC phase transition}
\begin{figure}
\begin{center}
  \includegraphics[width=10cm]{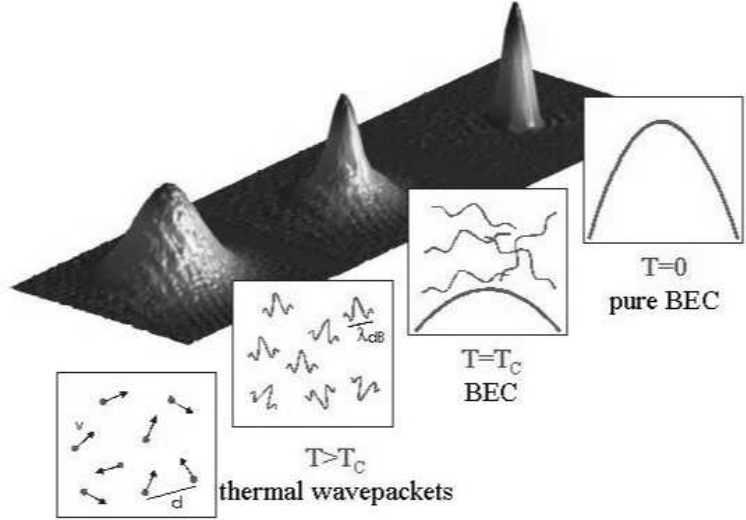}
\end{center}
  \caption{Simplified intuitive view of the Bose-Einstein phase
  transition in terms of a matter wave interpretation. The model
  system in the lower schematic images assumes constant particle number. The
  temperature decreases from the bottom left to the upper right,
  which leads to an increasing deBroglie wavelength of the
  particles. The phase transition occurs, when this wavelength
  becomes comparable to the mean particle distance.
  The upper images are experimental absorption images of an ensemble
  of $^{87}$Rb atoms at corresponding temperatures. These pseudo 3d images
  show the density distribution of the ensemble
  after a time of flight period. They thus show the corresponding momentum
  distribution of the cloud. The images visualise the transition from
  a thermal cloud with a symmetric momentum distribution to a pure quantum
  mechanical wavefunction with an asymmetric momentum
  distribution, reflecting the asymmetry of the trap used in the experiment.
  Below but still close to the BEC phase transition the system shows a
  two-component distribution corresponding to the normal
  fraction and the condensate wavefunction.}\label{fig:BECanschaulich}
\end{figure}

As already mentioned the phenomenon of Bose-Einstein condensation
directly follows from quantum statistics. It is thus based on very
few fundamental principles of quantum physics, mainly on the
indistinguishability of identical particles. Relying on the wave
aspect in the quantum mechanical description of particles, one can
also find an intuitive analogy to laser physics: We consider the
temperature dependend thermal de Broglie wavelength
\begin{equation}\label{deBroglie}
 \Lambda_{\rm dB}(T)=\sqrt{\frac{2\pi \hbar^2}{mk_BT}},
\end{equation}
with Planck's constant, $\hbar $, the particle mass, $m$,
Boltzmann's constant, $k_B$, and the ensemble temperature, $T$.
From this expression it becomes clear, that the particle
wavelength, which could be thought of as being on the same order
as the particle wave packet size, will grow with decreasing
temperature. Decreasing the temperature or increasing the density,
$n$, of an atomic ensemble will thus at some point lead to a wave
packet overlap, when the so called phase space density
\begin{equation}\label{PhaseSpaceDensity}
 n\Lambda_{\rm dB}^3(T)
\end{equation}
becomes of order unity. Taking further into account
Bose-enhancement of stimulated emission into occupied modes,
scattering-based redistribution processes of the deBroglie waves
will result in a non-classical equilibrium distribution. The
ground mode of the system, typically having the largest
occupation, will benefit most from this redistribution and the
nonlinearity of Bose-enhancement results in a macroscopic
occupation of this state. Fig.~\ref{fig:BECanschaulich} gives a
schematic view of this process. This intuitve picture is
impressively reflected in the formation process of a Bose-Einstein
condensate~\cite{Miesner1998b} as well as in matter wave
amplification~\cite{Inouye1999b} discussed in
section~\ref{s:matterwaveamplification}.

\begin{figure}
\begin{center}
\includegraphics[width=10cm]{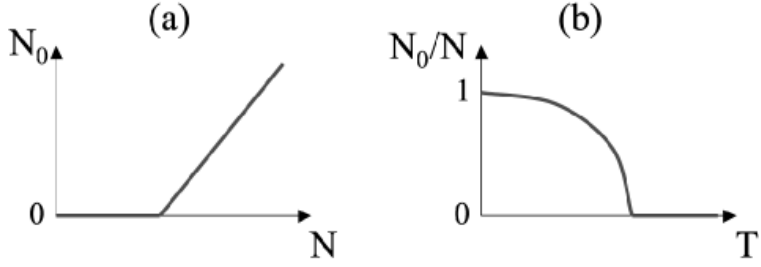}
\end{center}
     \caption{Schematic view of the dependence of ground state occupation number
     on the number of particles for a fixed temperature (a) as compared to the relative
     ground state occupation as a function of temperature for fixed particle number (b).}
     \label{fig:N0vonNvsT}
\end{figure}

A quantitative derivation of the Bose-Einstein condensation
phenomenom is obtained from the quantum statistical point of view.
Treating an ideal gas system in the convenient framework of the
grand canonical ensemble, the mean occupation number, $N_i$, of a
state, $i$, with energy $\epsilon_i$ of the system is given by the
Bose-distribution:
\begin{equation}
  N_i=\frac{1}{e^{(\epsilon_i-\mu )/k_BT}-1}.
\end{equation}
In this description the parameters temperature, $T$, and chemical
potential, $\mu $ fix the energy and particle number of the system
via:
\begin{eqnarray}\label{e:NandE}
 E &=& \sum_i \epsilon_i N_i=\sum_{i=0}^{\infty } \frac{\epsilon_i}{e^{(\epsilon_i-\mu
 )/k_BT}-1}\\
 N &=& \sum_i N_i=\sum_{i=0}^{\infty } \frac{1}{e^{(\epsilon_i-\mu
 )/k_BT}-1}.
\end{eqnarray}
For a given temperature the chemical potential defines the
occupation number of each energy level, which increases for
increasing $\mu $. Note that in order to avoid negative (and thus
unphysical) occupation numbers the chemical potential always has
to stay below the ground state energy, e.g. $\mu <\epsilon_0 $.
For a given temperature this effect leads to a maximum possible
number of particles in excited states
\begin{equation}\label{e:Nex}
  N_{\rm ex}=\sum_{i=1}^{\infty }\frac{1}{e^{(\epsilon_i-\mu
  )/k_BT}-1}
\end{equation}
for $\mu =\epsilon_0$. This number, $N_{\rm ex}(\mu = \epsilon_0)$
is finite for systems in which Bose-Einstein condensation occurs.
It is nevertheless possible to host more particles in these
systems as the ground state occupation diverges in the limit $\mu
\rightarrow \epsilon_0 $. Adding particles to such a system at
fixed temperature first enlarges $N_{\rm ex}$ by increasing $\mu $
until its value becomes close to $\epsilon_0 $. At this point the
excited state population saturates and the ground state becomes
macroscopically occupied (see Fig.~\ref{fig:N0vonNvsT} (a)), which
is the essence of Bose-Einstein condensation.

A slighly different but equivalent common viewpoint closely
related to the typical experimental realisation of Bose-Einstein
condensation considers cooling of a system with fixed particle
number. In this view the maximum possible excited state ocupation
decreases with decreasing temperature. Again at the point where
$N_{\rm ex}(\mu = \epsilon_0,T)$ becomes smaller than $N$, a
macroscopic part of the particles starts to occupy the ground
state (see Fig.\ref{fig:N0vonNvsT} (b)).

The critical temperature, $T_c$, and the corresponding critical
particle number, $N_c$, for Bose-Einstein condensation can be
found by considering the condition
 \begin{equation}
 N_{\rm ex}(\mu = \epsilon_0,T_c)=N_c.
 \end{equation}

A more quantitative analysis for systems in a relatively general
potential, $U(\vec r)$, can be obtained approximating the sum in
equation~\ref{e:NandE} for large particle numbers by the following
integral \cite{Goldman1981a,Huse1982a,Bagnato1987a}:
\begin{equation}\label{continousN}
 N=N_0+\int_0^{\infty }\rho (\epsilon ) N(\epsilon ).
\end{equation}
For this expression the ground state energy was set to be zero and
the continous Bose distribution
\begin{equation}
N(\epsilon )=\frac{1}{e^{(\epsilon - \mu -)/k_BT}-1}
\end{equation}
as well as the energy density of states
\begin{equation}
\rho_{\epsilon} = \frac{1}{2\pi \hbar^3}\int \int d^3r d^3p \,
\delta \left( \epsilon - \frac{p^2}{2m}-U(\vec r)\right)
\end{equation}
were introduced. Performing the momentum integration and comparing
with $N=\int d^3r\, n(\vec r)$ yields the particle density
\begin{equation}
n(\vec r)=\frac{g_{\frac{3}{2}}\left( e^{-(U(\vec r)-\mu
  )/k_BT}\right)}{\Lambda_{\rm dB}^3(T)}
\end{equation}
introducing the Bose-function $g_{\alpha }(x)=\sum_{l=1}^{\infty
}\frac{x^l}{l^{\alpha }}$. The critical temperature immediately
follows by considering $\mu \approx 0$ (assuming the minimum of
the potential $U(\vec r_0)$ to be zero) from
\begin{equation}\label{BECcondition}
  n(r_0)\cdot \Lambda_{\rm dB}^3(T_c)=g_{\frac{3}{2}}(1)\approx
  2.612
\end{equation}
The product of density and the cube of the de Broglie wavelength
ocurring in this equation is called the phase space density
\begin{equation}
D(T,n)=n\cdot \Lambda_{\rm dB}^3(T)
\end{equation}
of the ensemble and represents a central parameter in the
description of Bose-Einstein condensation. Above $T_c$ its cubed
root can be interpreted as the ratio of thermal de Broglie
wavelength and mean interparticle spacing with the above condition
corresponding to the intuitive picture given at the beginning of
this section. Note, that the condition~\ref{BECcondition} is
independend of the potential (as long as it has a minimum, with
the box potential as limiting case).

For the experimentally relevant case of an anisotropic harmonic
trap with frequencies $\omega_x $, $\omega_y $ and $\omega_z $ one
can extract a condition involving critical temperature and
particle number
\begin{equation}\label{Tcharmonic}
  k_BT_c\approx 0.94 \hbar (N_c\omega_x \omega_y
  \omega_z)^{\frac{1}{3}}.
\end{equation}

The above equations for the critical temperature are strictly
valid only for noninteracting gases in the thermodynamic limit of
particle numbers going to infinity. In the experimentally
important case of weakly interacting dilute atomic gases with
finite particle number (typically $N=10^3..10^7$) corrections on
the order of a few percent appear (see
e.g.~\cite{Pathria1998a,Dalfovo1999a}). These corrections are
observable in experiment and an active area of
research~\cite{Ensher1996a,Mewes1996a,Gerbier2004a}. Furthermore
in the regime of finite particle number differences between
different thermodynamic treatments occur, in particular concerning
fluctuations. Detailed descriptions via the most appropriate
microcanonic ensemble and corrections to it are an active area of
research~\cite{Gaijda1997a,Navez1997a,Wilkens1997a,Balazs1998a,Holthaus1998a,Idziaszek1999a,Borrmann1999a,Kocharovsky2000a,Kocharovsky2000b,Kocharovsky2000c,Holthaus2002a,Xiong2002a}.

It is important to note, that the critical temperature is usually
several orders of magnitude higher than the temperature
corresponding to the energy level spacing between the ground and
first excited state of the system. In this sense Bose-Einstein
condensation has nothing to do with naturally freezing out the
atomic motion but really is a high temperature phenomenon. This
availability of a single macroscopically occupied wave function is
one of the experimentally most intriguing aspects of Bose-Einstein
condensation as it gives access to coherent matter waves with
large coherence length. We will discuss role of finite temperature
effects concerning the coherence properties later in this article.

\subsubsection{Condensate wave function - Gross Pitaevskii equation}
The Bose-Einstein condensation phase transition discussed above is
only weakly influenced by interactions. For temperatures above the
critical temperature the interaction energy of the system is small
compared to the kinetic energy of the system. The Bose-Einstein
condensation phase transition itself is not significantly
influenced by interactions. On the other hand for temperatures
below the critical temperature, the condensate wavefunction, which
is of main interest to this report, is significantly altered.
Interparticle interactions significantly change the shape and size
of the condensate wavefunction but also introduce collective
excitations such as oscillations and sound as well as nonlinear
effects. For the condensate fraction the kinetic energy is usually
much smaller than the interaction energy (so called ''Thomas Fermi
regime''). A good overview of the condensate wavefunction theory
can be found in~\cite{Dalfovo1999a} on which the following
introduction of main concepts and approximations is based. In the
Heisenberg picture in second quantization, the time evolution of a
dilute gas of bosonic atoms can be expressed as
\begin{equation}\label{feldopgl}
  \fl i\hbar\frac{\partial }{\partial t}\hat \Psi^{\dagger}(\vec{r},t)=\left[
  -\frac{\hbar^2\nabla^2}{2m}+U_{\rm ext}(\vec r) + \int d\vec{r'} \hat
  \Psi^{\dagger}(\vec{r'},t)U(\vec r-\vec{r'})\hat
  \Psi(\vec{r',t})\right]\hat \Psi^{\dagger}(\vec{r'},t).
\end{equation}
$\hat \Psi^{\dagger}(\vec r)$ ($\hat \Psi(\vec r)$) is the boson
field operator creating (annihilating) a particle at position
$\vec r$ and $U(\vec r-\vec{r'})$ is the two particle interaction
potential. Solving the corresponding many body problem is
nontrivial and usually involves elaborate numerical simulations.

A very useful approximation for the case of low temperatures
$T\approx 0$ is to replace the bosonic field operators in lowest
order by their expectation value
\begin{equation}\label{Phidef}
  \Phi(\vec r,t)\equiv \langle \hat \Psi(\vec{r},t) \rangle ,
\end{equation}
introducing the condensate wave function, $\Phi(\vec r,t)$. The
introduction of a well defined phase as associated with this
condensate wave function is relevant to matter wave interference
experiments discussed below. It corresponds to the assumption of a
spontaneously broken symmetry in connection with the Bose-Einstein
phase transition.

Using the above definition in equation~\ref{feldopgl} results in
the Gross Pitaevskii equation
\begin{equation}\label{GPT}
  i\hbar \frac{\partial }{\partial t}\Phi(\vec r,t)=\left( -\frac{\hbar^2\nabla^2}{2m}+U_{\rm ext}(\vec
  r)+g|\Phi(\vec r,t)|^2\right) \Phi(\vec r,t),
\end{equation}
the time independend version of which is obtained using the ansatz
$\Phi(\vec r,t)=\phi(\vec r)\exp (-i\mu t/\hbar )$:
\begin{equation}\label{GP}
  \left( -\frac{\hbar^2\nabla^2}{2m}+U_{\rm ext}(\vec r)+g|\phi(\vec r)|^2\right) \phi(\vec
  r)=\mu \phi(\vec r).
\end{equation}
In these equations, $\mu $, is the chemical potential and
\begin{equation}
  g=\frac{4\pi \hbar^2 a}{m}
\end{equation}
is the coupling constant representing the interparticle
interactions. The parameter $a$ is the so called s-wave scattering
length. The approximation that the interactions are described by
s-wave scattering only is very well fulfilled for the low
temperature regime in typical experiments. $\phi(\vec r)$ can be
chosen to be a real function with $N=N_0=\int d\vec r \phi^2$ and
thus related to the particle density as $\phi^2(\vec r)=n(\vec
r)$.

In the low temperature regime the Gross Pitaevskii
equation~\cite{Gross1963a,Pitaevskii1961a} is the main
approximation for the condensate wave function. Is has been shown
in many experiments to be valid up to a high precision for the
case of weakly interacting atomic gases discussed in this report.
The Gross Pitaevskii equation not only very well describes the
ground state wave function, but also many other effects such as
collecive excitations, the condensate expansion, interference
properties, sound propagation, solitons and vortices.

It is intuitive to consider the energy contributions to a trapped
atomic Bose-Einstein condensate in order to better understand the
properties of the condensate wavefunction:
\begin{equation}
 \fl E[n]=\int d\vec r \left[ \frac{\hbar^2}{2m}|\nabla
  \sqrt{n}|^2+nU_{\rm ext}(\vec r)+\frac{gn^2}{2}\right] = E_{\rm
  kin}+E_{\rm ho}+E_{\rm int}.
\end{equation}
$E_{\rm kin}$ is the kinetic energy and can be interpreted as a
''quantum pressure'', $E_{\rm ho}$ is the potential energy due to
the trapping potential and $E_{\rm int}$ is the mean field
interaction potential energy of the atoms.

The kinetic and interaction energy terms determine the scale over
which density variations of the condensate wavefunction can occur.
This scale is given by the so called healing length and stems from
the fact that the condensate density can not grow from $0$ to $n$
on arbitrarily short distances as this would imply a divergence of
the ''quantum pressure'' term, $E_{\rm kin}$.  The shortest
possible distance, $\xi $ for such a change in density is given by
the balance of $E_{\rm kin}\sim \hbar^2/(2m\xi^2)$ and interaction
energy $E_{\rm int}\sim 4\pi \hbar^2 a n/m$ which defines the
healing length
\begin{equation}\label{e:healinglength}
  \xi = \frac{1}{\sqrt{8\pi n a}}.
\end{equation}
For $^{87}$Rb and a density of $n=10^{14}$ this length is $\xi
\approx 0.3\, \mu $m. A singularity as in the center of a vortex
or a matter wave soliton thus have a dimension of at least the
healing length.

The Gross Pitaevskii equation can be extended to include small
amplitude collective
excitations~\cite{Pitaevskii1961a,Fetter1972a} using the
Bogoliubov approximation. This extension can be derived by writing
the atom field operator from equation~\ref{feldopgl} as
\begin{equation}
  \hat \Psi (\vec r,t)=\Phi (\vec r,t) + \hat \Psi '(\vec r,t).
\end{equation}
$\Phi(\vec r,t)$ is the ground state wave function defined in
equation~\ref{Phidef} and
\begin{equation}
  \Psi '(\vec r,t)=\sum_j{\left(u_j(\vec r)\alpha_j (t)+v_j^*(\vec
  r)\alpha_j^+(t) \right) }
\end{equation}
represents excitations. In this equation $\alpha_j (t)$ and
$\alpha_j^+(t)$ are Bogoliubov quasiparticle annihilation and
creation operators with the respective amplitude functions
$u_j(\vec r)$ and $v_j^*(\vec r)$. In a uniform gas $u$ and $v$
can be represented as plane waves obeying the Bogoliubov
dispersion law~\cite{Bogoliubov1947a}
\begin{equation}
  \hbar \omega = \sqrt{\frac{\hbar^2q^2}{2m}\left(
  \frac{\hbar^2q^2}{2m}+2gn\right) },
\end{equation}
with the excitation wavevector, $\vec q$. In the case of large
momenta the excitations behave as free particles with energies
$\hbar^2q^2/2m$. Low momenta yield a phonon dispersion law $\omega
= cq$ with the sound velocity of the system
\begin{equation}\label{e:soundvelocity}
  c=\sqrt{\frac{gn}{m}}.
\end{equation}
For typical experimental parameters the sound velocity is on
the order of a few $\frac{\rm mm}{\rm s}$.

\subsubsection{Beyond the Gross Pitaevskii equation}

In principle the Gross Pitaevskii equation only describes the
physics of weakly interacting systems for $T\rightarrow 0$, i.e.
the pure condensate fraction. Nevertheless the Gross Pitaevskii
equation has shown to give an excellent qualitative and
quantitative description for nearly all experiments to date.
Finite temperature effects are key to important experimental
effects such as damping of excitations and especially decoherence
phenomena like phase fluctuations. The theory can be very
involved, and suitable approximations are an active area of
research. Promising theoretical routes for temperatures close to
and below $T_c$ consist in extensions of the Gross Pitaevskii
equation, modifying the well known pure condensate fraction with
an additional
reservoir~\cite{Proukakis1996a,Gardiner1997a,Gardiner1997b,Hutchinson1997a,Gardiner1998a,Gardiner1998b,Choi1998a,Rusch1999a,Zaremba1999a,Gardiner2000a,Goral2002a}.

For the case of sufficiently strong interactions, e.g. for
Bose-Einstein condensates in strong periodic potentials, the
system becomes strongly correlated. The physics of such a system
goes fully beyond the Gross Pitaevskii equation and demands for
different theoretical descriptions. In this regime effects like
large phase uncertainty due to number squeezing~\cite{Orzel2001a}
and the superfluid to Mott insulator transition can
occur~\cite{Greiner2002a}.

\section{Experimental realisation of BEC}
\begin{figure}
\begin{center}
\includegraphics[width=12cm]{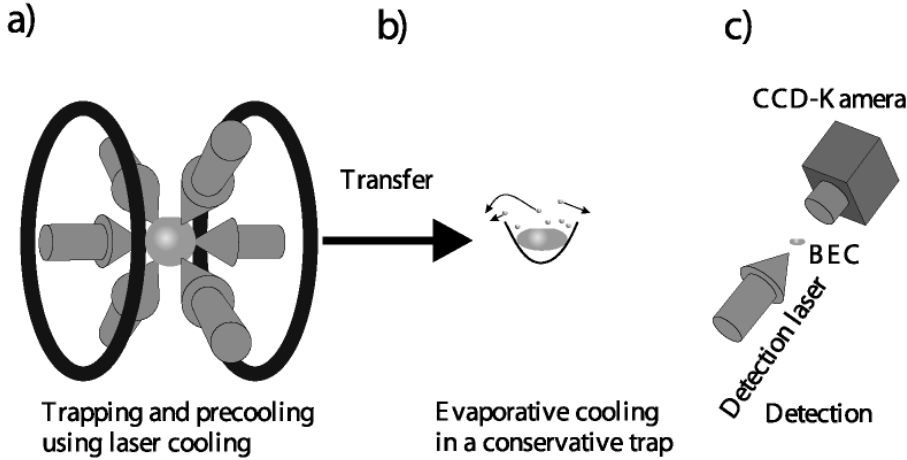}
     \caption{Typical setup for the creation of a Bose-Einstein condensate in dilute alkali gases.
                 Note that the steps a)-c) are taking place sequentially in time but at the
                 same position in space. First atoms are collected
                 and precooled in a magnetooptic trap (MOT)(a). In order to avoid photon induced heating
                 processes the ensemble is then transferred to a conservative potential trap
                 (here: a magnetic trap). Further cooling and
                 compression in the conservative trap is done by evaporative cooling, which removes hot
                 atoms from the ensemble until the BEC phase transition is reached (b).
                 At all stages of the experiment the ensemble can be observed by
                 direct imaging onto a CCD chip with absorption or
                 dispersive techniques (c).}
     \label{fig:BECErzeugung}
\end{center}
\end{figure}

In this section we will give a short summary of the experimental
realization of Bose-Einstein condensation in dilute atomic gases.
Detailed descriptions can be found in several review articles,
e.g.~\cite{Ketterle1999a,Cornell1999a}. The typical experimental
approach consists in cooling a magnetically or optically trapped
ensemble of atoms in an ultra high vacuum ($\approx 10^{-11}\,
$mbar) environment, which provides thermal isolation. For alkali
atoms cooling is typically a two step process. First
$10^8-10^{11}$ atoms are collected and precooled to $10-100\mu $K
at densities around $10^{10}-10^{11}$ cm$^{-3}$ using laser
cooling techniques. This step increases the phase space density
from around $10^{-12}$ to $D\approx 10^{-6}$. Then the ensemble is
transferred into a conservative potential and further cooled and
compressed with evaporative cooling until the Bose-Einstein phase
transition ($D=2.612...$) occurs. This point is typically reached
with $10^4-10^7$ atoms at $100\,$nK-$1\, \mu$K and densities
around $10^{14}\,$cm$^{-3}$. The whole experimental cycle
typically takes between 10 and 100$\, $s. These steps are
summarised in Fig.~\ref{fig:BECErzeugung}.

It is interesting to note, that this typical experimental approach
to reach Bose-Einstein condensation via cooling is ''orthogonal''
to the original idea of Einstein~\cite{Einstein1925a}. Einstein
suggested an isothermal compression of the system, which for
trapped atomic ensembles would correspond to an isothermal
increase of particle number (see Fig.~\ref{fig:N0vonNvsT} (a)).
From this viewpoint the condensation is very intutive, as the new
particles first all add to the normal component (the so called
''thermal cloud'') until it reaches its saturated occupation for
the given temperature after which practically all further
particles exclusively add to the condensate wavefunction. The
constant temperature approach was recently realised in an
experiment~\cite{Erhard2004a} using spinor condensates.

\section{Coherence properties of BEC}
The occurrence of a macroscopically occupied ground state
wavefunction, i.e. a ''single mode'' coherent matter wave, is one
of the most intriguing aspects of Bose-Einstein condensation. The
experimental observation of Bose-Einstein condensation in dilute
alkali gases initiated many research activities concerning the
''physics with coherent matter waves''. The coherence was
impressingly demonstrated in pioneering interference
experiments~\cite{Andrews1997b,Burt1997a,Hall1998b} short after
the first experimental realisation of Bose-Einstein condensation
in dilute atomic gases. This research covers fundamental aspects
of quantum coherence as well as promising applications in the
field of high precision atom interferometry~\cite{Berman1997a}.
The most outstanding prospect connected to Bose-Einstein
condensation as source for coherent matter waves lies in the
analogy to the revolution of light optics with the introduction of
laser sources.

A theoretical description of the coherence properties of dilute
Bose gases based on the theory of optical coherence can e.g. be
found in~\cite{Naraschewski1999a}. In the following we will
discuss the coherence properties of a Bose-Einstein condensate in
view of experimental possibilities and limitations.

\subsection{First order coherence}
In a simple picture first order coherence is intrinsically related
to the visibility of interference fringes, when two waves overlap.
Furthermore the coherence length defines the maximum relative
position shift two interfering wavepackets can have while still
displaying interferences. These aspects are of major importance
for the implementation of interferometric schemes and lead to
several fundamental investigations and fascinating experiments.

The coherence of the Bose-Einstein condensate wavefunction as a
macroscopic matter wave and its capability to interfere were some
of the most intriguing questions after the first realisation of
Bose-Einstein condensation in dilute atomic gases. In order to
understand the controversies over this subject, despite the
theoretical prediction of an order parameter with well defined
phase (see eq.~(\ref{Phidef})), one has to consider the
experimental realisation. While the theoretical prediction is
based on the thermodynamic limit, i.e. infinite particle number,
the experiments were performed with dilute atomic gases, reaching
Bose-Einstein condensation with typically $10^3..10^7$ atoms. As
indicated above, in this case the thermodynamic ensembles for the
statistical treatment are not equivalent. The most appropriate
description of the experiment seems to be via the microcanonical
ensemble, dealing with a given finite particle number with given
energy in a trap. Fixing the particle number directly in the
microcanonical ensemble in comparison to ensuring a statistically
most probable particle number via the chemical potential in the
grand canonical ensemble would suggest a despription of the
condensate wavefunction as a Fock state rather than a coherent
state. Due to the number phase uncertainty relation a Fock state
with well defined particle number has maximum phase uncertainty.
The fundamental controvery of major importance concerning the
matter wave coherence was if two seperately created condensates
would be able to form a visible interference pattern and if this
would be consistent with a Fock state description. This controvery
was resolved by the experiment described below and the theoretical
prediction, that two Fock states can indeed form an interference
pattern, when
overlapped~\cite{Naraschewski1996a,Wong1996a,Javanainen1996c,Hoston1996a,Rohrl1997a,Castin1997b,Javanainen1997a}.

It is important to note, that this effect relies on the
availability of a macroscopically occupied wavefunction. Due to
macroscopic occupation a random relative phase between the two
interfering Fock states can be detected in a single shot
experiment. For the average over many single particle experiments
the randomness of the relative phase washes out any interference
visibility. In this case there is a fundamental difference between
a single experiment with many particles and many experiments with
single particles!

A very intuitive approach to explain the relatice phase in Fock
state interference experiments was developed
in~\cite{Castin1997b}. The relative phase is built up by Bose
enhancement effects during the detection process. If some out of
many bosons occupying the same wavefunction are detected
(localised on some screen position) Bose enhancement leads to a
modification of the detection probability of further bosons. Due
to the position resolved detection the Fock state is projected
onto a state with well defined phase. This phase is expected to be
random for different experiments but fixed for each single
experiment. Note that there is also no contradiction between well
defined particle number and phase at the same time, as the
detection introduces a local number uncertainty for each screen
position.

\subsection{Interference experiments}\label{s:interferece}
The groundbreaking experiment establishing Bose-Einstein
condensates in dilute atomic gases as coherent matter waves was
performed by the group of W. Ketterle at MIT~\cite{Andrews1997b}.
This experiment established the interference capability of
Bose-Einstein condensates in a twofold way. It demonstrated self
interference (similar to the interference in typical light
interference schemes performed with one laser) as well as the
interference of two distinct condensate wavefunctions (similar to
''beat signal interferences'' between two distinct lasers).

\begin{figure}
\begin{center}
\includegraphics[width=8cm]{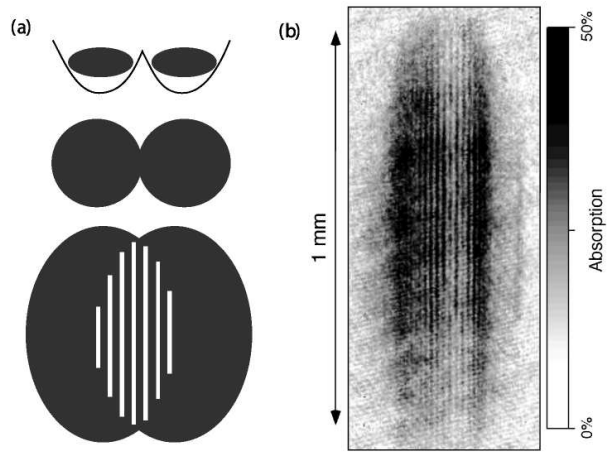}
     \caption{(a) Schematic view of the interference of two seperately
     produced Bose-Einstein condensates. After switching off the double
     well trapping potentials the condensate wavefunctions fall in gravity and
     expand. When they start to overlap interference fringes appear in the
     overlapping region. (b) Absorption images on BEC interference obtained
     in the group of W. Ketterle at MIT. Experimental image with courtesy of W. Ketterle, MIT.}
     \label{fig:BECInterferenz}
\end{center}
\end{figure}

This experiment made use of a trap for the atomic ensemble, which
could be devided into two seperate traps with an adjustable
repulsive potential well created with a light sheet. Switching off
the trapping potential with a Bose-Einstein condensate in each
trap causes the condensate wavefunctions to spread during time of
flight. When these wavefunctions spread into each other they show
interference fringes according to their coherence properties in
the overlap region (see Fig.~\ref{fig:BECInterferenz}).

With this setup a self interference experiment was performed by
first creating a Bose-Einstein condensate in the undivided trap
before slowly splitting the trap and thus also the condensate in
two parts. In this case (and neglecting decoherence effects) there
should be no distinct condensate wavefunctions, but only one with
a double peak distribution. Each atom should be in a superposition
state of both trap wavefunctions. There should thus be an
interference pattern after the time of flight expansion due to
interference of ''each atom with itself'' with a fixed phase only
depending on the splitting process.

The second experiment performed with this setup started with a
split trap and the creation of two distinct Bose-Einstein
condensates from a thermal cloud in each trap. In this case an
interference pattern would result from two individual matter waves
and the experiment should correspond to the controversy discussed
above. Note that now the relative phase of subsequent experiments
is expected to have a random distribution, shifting the position
of the interference pattern with each experimental run.

In both cases the experiments showed a clearly visible
interference pattern, confirming the first order coherence of the
Bose-Einstein condensate wavefunction for dilute atomic gas
experiments. Due to the technical difficulty of these experiments,
a distinction between a fixed and arbitrarily moving interference
pattern for the two preparation cases could not yet be observed.

The coherence of the condensate wavefunction in connection with
its macroscopic size open a new window to quantum mechanics. It
becomes possible to directly observe textbook examples for
interferometry~\cite{Andrews1997b,Burt1997a,Hall1998b,Kozuma1999a,Bongs1999a,Torii2000a,Minardi2001a,Pedri2001a}
and for the propagation of wavefunctions in experiment and use
them to investigate atom optical elements. An easy realisation is
already the reflection of a matter wave by a potential, e.g. an
atom mirror~\cite{Bongs1999a}. In this case self interference
effects occur, similar to the case of numerically propagating a
Schr\"odinger wavepacket towards a potential step.

\subsection{BEC as minimum uncertainty
state}\label{s:minimumuncertainty} One important issue concerning
the coherence of a Bose-Einstein condensate wavefunction is
whether its coherence length is equal to its size or shorter. From
another viewpoint this corresponds to the question, if the
wavefunction of a trapped Bose-Einstein condensate represents a
minimum uncertainty state with respect to the Heisenberg position
momentum uncertainty relation.

The spatial distribution of the condensate wavefunction can be
deduced from experiment by the standard technique of absorption
imaging in a straightforward way. The momentum distribution of the
trapped sample is however more difficult to obtain, as ordinary
time of flight measurements (converting the momentum distribution
into spatial distribution) are strongly influenced by the release
of mean field energy, the nonlinear term in the Gross Pitaevskii
equation~\ref{GP}. The first measurements of the condensate
momentum spread implemented Bragg spectroscopy as a new
tool~\cite{Kozuma1999a,Stenger1999b} in BEC experiments. Bragg
spectroscopy is a special case of Bragg
scattering~\cite{Martin1988a}, which has become a versatile
instrument for the manipulation and investigation of the
condensate wavefunction.

\begin{figure}
\begin{center}
\includegraphics[width=12cm]{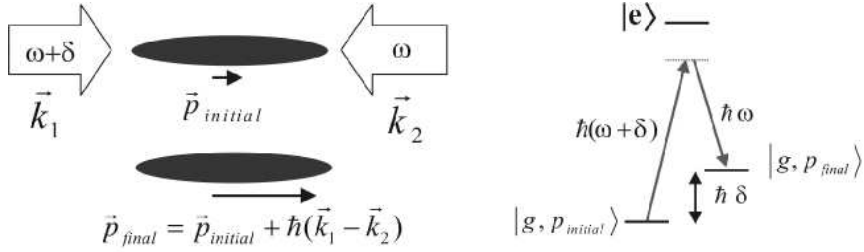}
\end{center}
     \caption{Scheme for Bragg diffraction as a Raman process.}
     \label{fig:Braggscheme}
\end{figure}

In principle Bragg diffraction simply corresponds to the
scattering of a matter wave off a light intensity grating, similar
to the scattering of light waves off crystal planes in
crystallography. From an atomic physics viewpoint Bragg scattering
can be interpreted as a Raman process, giving intuitive insights
into the relevant parameters~\cite{Kozuma1999a}. In the
corresponding picture two momentum states $|g,\vec p_{\rm
initial}\rangle $ and $|g, \vec p_{\rm final}= \vec p_{\rm
initial}+\hbar (\vec k_1 - \vec k_2)\rangle $ connected with an
atomic ground state are coupled by a two-photon transition. The
coupling is typically induced by a pulse from two laser beams
having wave vectors $\vec k_1$ and $\vec k_2$. For the following
discussion we will specialise on the case of counterpropagating
laser beams as depicted in Fig.~\ref{fig:Braggscheme}.

Fig.~\ref{fig:Braggscheme}

Imposing energy conservation for the above example of Bragg
diffraction with counterpropagating laser beams leads to the
relation
\begin{equation}\label{Braggcondition}
  \hbar \delta = E_{\rm final}-E_{\rm initial} = \frac{2}{m}(\hbar k p_{\rm initial } +
  \hbar^2k^2).
\end{equation}
For this Bragg condition the approximation $k_1=k_2=k$, the laser
frequency detuning, $\frac{\delta }{2\pi }$ and the particle mass,
$m$ were used. Energy conservation~(\ref{Braggcondition})
introduces a velocity dependence of Bragg scattering, enabling
momentum spectroscopy with this technique.

It is important to visualise, that the velocity selectivity
corresponds to the frequency width of the Fourier transform of the
temporal Bragg pulse envelope. As an example for $^{87}$Rb and
counterpropagating laser beams the velocity dependence of the
resonance condition is given by $\frac{\Delta v}{\delta /2\pi
}\approx 0.39\frac{\rm mm/s}{\rm kHz} $. In this sense there are
two regimes for Bragg diffaction employed on Bose-Einstein
condensates: 1. Bragg spectroscopy, with long pulses corresponding
to a higher velocity selectivity than the condensate velocity
spread~\cite{Kozuma1999a,Stenger1999b,Bongs2003a} and 2. Bragg
condensate optics, relying on short intense pulses, coupling an
entire condensate (with all its velocity classes) to a different
mean momentum state, e.g. to realise beamsplitters or
mirrors~\cite{Kozuma1999a,Hagley1999a,Hagley1999b,Torii2000a}.

\begin{figure}
\begin{center}
\includegraphics[width=12cm]{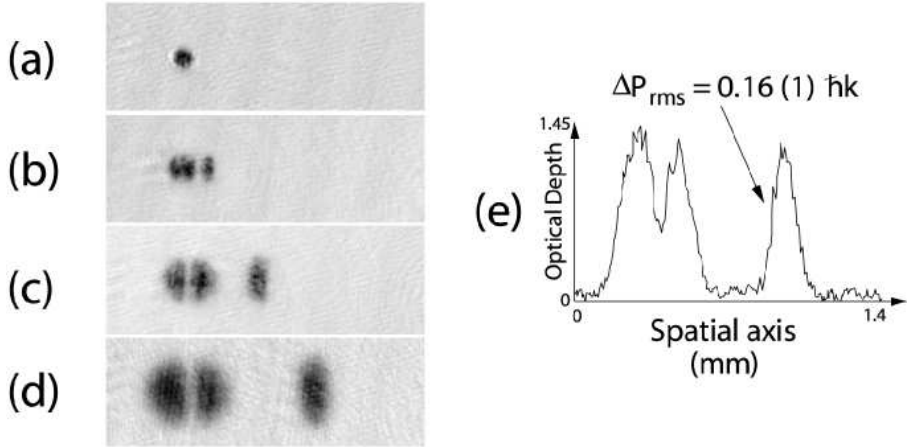}
\end{center}
     \caption{Bragg spectroscopy of a Bose-Einstein condensate. (a)-(d) show
     absorption images of the ensemble with an increasing time of flight after
     its interaction with the Bragg laser pulse. These images show the growth of
     the ensemble due to its momentum spread and the separation of the Bragg scattered
     momentum component (moving to the right). (e) shows the density distribution of
     the ensemble along the Bragg laser axis after separation of the scattered component
     from the remaining part of the condensate. Note that only a narrow momentum component,
     corresponding to a width of $0.16$ photon recoil momenta $\hbar k$, is scattered by
     the Bragg pulse. This graph clearly demonstrates the velocity
     selectivity of Bragg scattering in the long pulse regime. Images and graph with courtesy of
     the National Institute of Standards and Technology, USA.}
     \label{fig:Braggvsel}
\end{figure}

Bragg scattering is very useful for Bose-Einstein condensate
physics, as the associated momentum transfer is typically larger
than the momentum spread of the condensate. It consequently leads
to a time of flight separation of the atoms which have been
scattered from the remaining atoms by $\Delta x=\frac{2\hbar
kt_{\rm tof}}{m}$, where $t_{\rm tof}$ is the time of flight after
the Bragg pulse. Fig.~\ref{fig:Braggvsel} demonstrates this
splitting with an experimental example of high momentum resolution
Bragg spectroscopy.

Using the long pulse regime of Bragg spectroscopy results in the
transfer of a certain momentum component to a momentum state which
seperates from the condensate ensemble during time of flight. With
this process it is possible to extract the condensate wavefunction
momentum spread. A comparison with the expected value, based on
the spatial wavefunction spread and the minimum uncertainty given
by the Heisenberg uncertainty relation resulted in a good
agreement~\cite{Stenger1999b}.

\subsection{Phase measurements of BEC}
The predictability of the spatially varying phase of a
Bose-Einstein condensate is intrinsically related to its coherence
properties. For applications it is useful to know how the phase of
the condensate wavefunction evolves in different experimental
situations, e.g. during free expansion or inside a matter
waveguide. Phase measurements have to be relative measurements
with respect to a suitable reference object, as absolute phase is
not accessible in experiment. The phase evolution of the
condensate wavefunction can be studied in a well defined and
controlled way by the means of interferometric methods.

Here we discuss a method using a Bragg interferometer scheme (see
section~\ref{s:Bragginterferometer}), in which the condensate
phase evolution was measured during expansion in free
fall~\cite{Simsarian2000a} as well as in a matter
waveguide~\cite{Bongs2001a}. The experiments were based on open
interferometers, i.e. asymmetric interferometers in which the
wavefunction was split in a coherent superposition of two momentum
states, but then was subjected to a variable spatial overlap in
the output ports. The autocorrelation function of the condensate
wavefunction can be extracted by systematic variation of the
spatial displacement of the two interfering wavepackets in
addition to spatially resolved detection of the interference
pattern (see also Fig.~\ref{fig:Bragginterferometer}).

In the measurements of~\cite{Simsarian2000a} an equidistant fringe
pattern was found with a spatial fringe frequency varying with the
spatial overlap of the wavepackets. This can be explained within
the Thomas-Fermi approximation for the condensate wavefunction. In
this limit the ensemble is dominated by mean-field energy and the
phase of the trapped condensate wavefunction is constant. After
release from the trap the mean-field energy is converted to
kinetic energy, which for an initial harmonic trapping potential
leads to a quadratic phase profile for each
wavepacket~\cite{Kagan1996a,Castin1996a}. In addition mean field
energy leads to a repulsion of the wavepackets during the Bragg
beamsplitting process, which adds a relative velocity and thus a
linear phase term to the wave function. The phase of the wave
function $fe^{i\phi }$ in the direction of the splitting can thus
be assumed as~\cite{Simsarian2000a}
\begin{equation}
  \phi = \frac{\alpha }{2} x^2 + \beta x,
\end{equation}
with the spatial coordinate $x$ and $\alpha $ and $\beta $ as
parameters. The interference pattern originates from the two
overlapping wavepackets at the last beamsplitter $f(x-\delta
x)e^{i\phi (x-\delta x)}$ and $f(x)e^{i\phi (x)}$, with the
spatial distance of the wavepacket centers, $\delta x$. The
resulting fringes are predicted to be equidistant with a spatial
frequency~\cite{Simsarian2000a},
\begin{equation}
  \kappa = \alpha \delta x + 2\beta .
\end{equation}

The above experiments clearly showed the coherence of the
Bose-Einstein condensate wavefunction in important experimental
situations with a coherence length comparable to its extends.
Furthermore the finding of an equadistant fringe spacing confirmed
the Thomas-Fermi mean field approximation theory to provide an
accurate description of the coherent matter wave dynamics.

\subsection{Phase fluctuations}\label{s:phasefluct}
The conclusion that a trapped Bose-Einstein condensate has a
coherence length corresponding to its extends is not true in
general. It is only valid fo the case of nearly spherical
condensates and very low temperatures $T\ll T_c$. For the case of
very elongated trapping geometries and temperatures comparable to
$T_c$ important discrepancies occur.

In this case the excitations connected to finite temperature cause
phase fluctuations. Also concerning the dynamics of the formation
process of a condensate wavefunction the condensation process can
be assumed to occur in two steps, first establishing the density
equlibrium distribution and only later full phase coherence. The
ensemble thus shows temporal phase
fluctuations~\cite{Kagan1992a,Kagan1994a,Kagan1995a,vanDruten1997a}
during the phase transition.

The character of Bose-Einstein condensation strongly depends on
the dimension of the system. In 1D even no Bose-Einstein phase
transition occurs but regimes like the Tonks gas appear. Thus one
can understand that deforming a spherical condensate grometry
towards 1D will already change the physics.

Here we focus on the equilibrium state of a low-dimensional (1D
and 2D) quantum gas
\cite{vanDruten1997a,Monien1998a,Olshanii1998a,Petrov2000a,Petrov2000b,Kagan2000a,Andersen2002a,Khawaja2002a}
which is expected to show persistent phase fluctuations and
similar effects in finite temperature condensates in very
elongated 3D trapping geometries \cite{Petrov2001a}. These
fluctuations are connected to thermal excitations with energies
below the lowest radial excitation energies but high enough to
cause axial excitations. In a handwaving picture one could argue,
that information on fluctuations on one end of an elongated
condensate wavefunction (e.g. due to interaction with atoms from
the normal component) cannot travel to the other end on a
timescale smaller than the fluctuation to occur. In corresponding
investigations it was indeed found that the phase coherence length
in this case can be much smaller than the axial size of the
sample~\cite{Dettmer2001a,Hellweg2001a,Shvarchuck2002a,Kreutzmann2003a,Hellweg2003a,Gerbier2003a,Mora2003a,Cacciapuoti2003a,Richard2003a,Khawaja2003a}

Theoretical predictions result in a temperature dependend phase
coherence length, $l_{\phi }$\~cite{Petrov2001a,Kreutzmann2002a},
with
\begin{equation}
R_{z}/l_{\phi}=\!16\left(\frac{a m^{1/2}}{15^{3/2}
\hbar^3}\right)^{2/5}\!\! \frac{k_{\mbox{\tiny
B}}T}{N_0^{3/5}}\left(\frac{\lambda}{\omega_z}\right)^{4/5}.
\end{equation}
In this equation $R_z=\sqrt{\frac{2\mu }{m\omega_z^2}}$ is the
Thomas Fermi axial radius of the condesate wavefunction,
$\lambda=\frac{\omega_{\rho }}{\omega_z}$ is the ratio of the
radial and axial trap frequencies and $N_0$ represents the number
of condensed atoms. The phase coherence length becomes smaller
than the axial size of the condensate below the characteristic
temperature~\cite{Petrov2001a}
\begin{equation}
  T_{\phi }=\frac{15 \hbar^2 \omega_z^2 N}{32\mu },
\end{equation}
with the chemical potential $\mu $.

\begin{figure}
\begin{center}
\includegraphics[width=6cm]{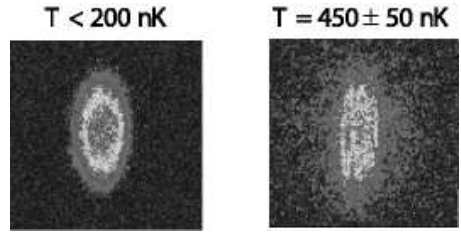}
\end{center}
     \caption{Absorption images of Bose-Einstein condensates produced in an
     elongated trap after 24.5\,ms time of flight. For low
     temperature (left)the ensemble exhibits long range phase coherence, while
     for temperatures close to $T_c$ (right) phase fluctuations occur (quasicondensate regime).
     The new momentum components due to phase fluctuations convert to density modulations
     after some time of flight, as visible in the right image. Images from the group of W. Ertmer, Hannover university.}
     \label{fig:phasefluct}
\end{figure}

One example for the experimental observation of phase fluctuations
is given by release measurements from the elongated trapping
potential as shown in Fig.~\ref{fig:phasefluct}. For phase
fluctuating Bose-Einstein condensates random striations in time of
flight measurements occur due to higher momentum components in the
condensate wavefunction.

Phase fluctuations are of central importance for many applications
of Bose-Einstein condensates. They pose severe limits on the
realisation on matter waveguides and are thus particularly
important for the rapidly evolving field of microscopic traps and
atom chips~\cite{Schlosser2001a,Folman2002a,Dumke2002a}.

\subsection{Second and third order coherence measurements}
Considering the comparison of light and atom optics, the question
weather the matter wavefunction created in the Bose-Einstein phase
transition also shows higher order coherence, is of particular
importance to distiguish ''atom laser'' like behaviour from
''just'' monochromatic matter waves. The link between higher order
loss processes and the according condensate
correlations~\cite{Burt1997a,Ketterle1997a} represented an
important step towards the measurement of higher order
correlations of the condesate wavefunction. Second and third order
correlations of the condensate wavefunction were determined by a
comparison of the measuring rate coefficients for two- and
three-body loss processes for the case of a thermal (i.e.
uncorrelated) gas and a Bose-Einstein condensate. The experiments
indeed found a two-fold reduction of the two-body loss
rate~\cite{Ketterle1997a} as well as a six-fold reduction of the
three body loss rate~\cite{Burt1997a} for the condensed sample.
These reduction factors clearly demonstrate second and third order
coherence of the condensate wavefunction and are also expected
from quantum statistics due to the indistinguishability of atoms
occupying the condensate wavefunction. Note, that this type of
measurements is restricted to small distance (on the order of
particle scattering lengths) correlations, as the typical
interatomic distance relevant to loss processes is much smaller
than the dimension of the condensate wavefunction.

Although second and third order correlations have been
investigated for the short distance regime higher order condensate
correlations on distances comparable to the size of the condensate
are relatively unexplored. There is however growing interest in
corresponding measurements, as they promise significantly expand
our knowledge on coherent matter waves. As one example higher
order correlation measurement techniques have been used to gain
detailed information on phase fluctuating Bose-Einstein
condensates~\cite{Cacciapuoti2003a}.

The field of large distance second and higher order correlations
of coherent matter waves is at the beginning of an interesting
development with still many open questions.

\section{Phase manipulation of BEC}
The large size of the Bose-Einstein condensate wavefunction opens
unique possibilities for the study of quantum mechanics. While
most experiments rely on the ''direct'' detection of this coherent
matter wave by optical imaging, the process of its ''direct''
local manipulation with suitable potentials or interactions seems
even more exciting. One possible realisation of this quantum
engineering makes use of destructive interactions, e.g. by the
controlled removal of atoms from the condensate with a resonant
laser beam. As an example this method has been used to increase
the average rotation in rapidly rotating condensates, by selective
removal of atoms from the low angular momentum centre of the
ensemble~\cite{Engels2003a}.

In this section we will introduce the concept of non-destructive
quantum engineering by phase manipulation of the condensate
wavefunction, in particular by phase imprinting.

\subsection{Concept of phase imprinting}
Phase imprinting relies on the (simplified) idea, that the local
phase evolution of a quantum mechanical wave function is governed
by its Hamiltonian evolution
\begin{equation}
  \psi(\vec r,t)\propto e^{\frac{i}{\hbar }H(\vec r,t)t}\psi(\vec r,0).
\end{equation}
In case $\psi $ is an Eigenfunction of the Hamiltonian and $E$ is
the corresponding Eigenenergy, this relation simplifies to a pure
phase change
\begin{equation}
  \psi(\vec r,t)\propto \psi(\vec r,0)e^{\frac{i}{\hbar }E(\vec r,t)t}.
\end{equation}
In experiment the Hamiltonian $H(\vec r,t)=H_0(\vec r,t)+V(\vec
r,t)$ can be controlled by adding an external potential $V(\vec
r,t)$ to the remaining energy terms contained in $H_0$. The
external potential can be produced in a convenient way with a
pulsed, spatially modulated and far off resonant laser beam,
leading to a light shift potential. The main idea is that the
additive potential creates an additional phase distribution along
the condensate wavefunction on a very short timescale, such that
other phase influences during the exposure are nearly constant and
negligible. The pure phase shifting action aimed at in experiment
can be approximately realised in this way, if the following is
fulfilled: the external potential is arranged to act on all
occupied Eigenfunctions of the Hamiltonian in the same way and in
addition it is much larger than the other energy terms, i.e $V\gg
H_0$. In this way the phase of the quantum mechanical wavefunction
can be significantly changed, by pulsing on a strong laser field
for a time much shorter than its correlation time, $\tau_c$. The
correlation time, $\tau_c=\frac{\xi }{c}$ is given by the time a
wave function disturbance moving with the condensate sound
velocity, $c$, (see Eq.~(\ref{e:soundvelocity})) over a distance
corresponding to the condensate healing length, $\xi $ (see
Eq.~(\ref{e:healinglength})).

The method of phase imprinting has been proposed with the main aim
to create circular flow with vortices in a Bose-Einstein
condensate~\cite{Dobrek1999a}. It was however already pointed out,
that this method is very general and that nearly arbitrary
manipulations of the wave function can be performed with it. This
is in particular true, if the method is extended by additional
resonant laser fields, allowing amplitude manipulations also, so
that amplitude and phase can be ''designed''. In principle full
control of the condensate wavefunction could be attained if the
optical resolution would be below the healing length scale.

Note that the method of phase imprinting is related to holographic
laser beam manipulation techniques in optics and represents a
similarly powerful tool for coherent matter wave optics.

\subsection{Experiments on phase manipulation of
BEC}\label{s:phasemanip} The intriguing possibility to
''engineer'' the quantum mechanical wavefunction of a
Bose-Einstein condensate using controlled phase manipulation
enables fascinating experiments. Phase manipulation with the help
of the earth gravitational potential or a similar acceleration
pseudopotential was employed to modify the interference pattern in
experiments with Bose-Einstein condensates in optical
lattices~\cite{Orzel2001a,Morsch2001a}. Furthermore the creation
of a new hyperfine state condensate component with a circular
$2\pi $ phase gradient by spatially phase modulated coupling with
an ordinary condensate resulted in the first experimental creation
of a Bose-Einstein condensate vortex state in dilute atomic
gases~\cite{Matthews1999b}.

\begin{figure}
\begin{center}
\includegraphics[width=12cm]{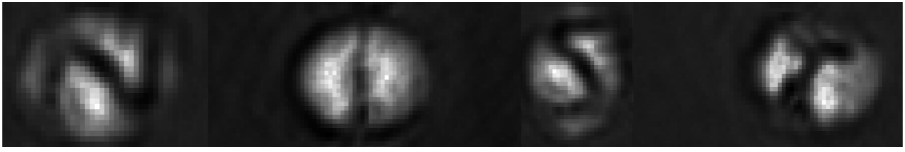}
\end{center}
     \caption{Experimental example of the possibilities of phase imprinting. The image is created
     from four different absorption images of condensates with different letters imprinted in the
     phase of the wavefunction. The images are taken in an interferometric setup which converts
     the phase information into density information (see~\cite{Denschlag2000a}). Images with courtesy of the
     National Institute of Standards and Technology (NIST), USA.}
     \label{f:NIST}
\end{figure}

Direct imprinting of a phase pattern with the help of suitably
shaped laser fields as discussed above has the advantage of being
very flexible. In principle all kinds phase distributions can be
created within the condensate wavefunction, e.g. the letters
''NIST'' as shown in Fig.~\ref{f:NIST}.

\begin{figure}
\begin{center}
\includegraphics[width=12cm]{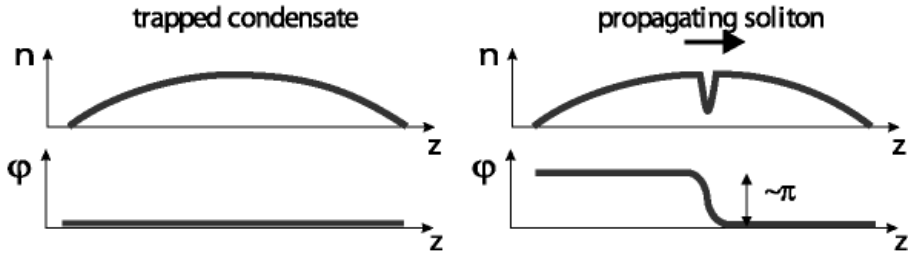}
\end{center}
     \caption{Schematic view of the density and phase distribution of a dark soliton
                 in a Bose-Einstein condensate wavefunction (right) as compared to the
                 condensate ground state (left).}
     \label{f:DSdist}
\end{figure}

In this section we will concentrate on the creation of dark
solitons as a first use of the method of phase
imprinting~\cite{Burger1999a,Denschlag2000a}. The physics of dark
solitons themselves will be addressed in more detail in
section~\ref{s:solitons}. In order to understand their creation
with the help of phase imprinting it is however important to
visualise their density and phase distribution. The schematic view
in Fig.~\ref{f:DSdist} shows that the density distribution of a
condensate with a dark soliton differs from the density
distribution of an unperturbed Bose-Einstein condensate
wavefunction only in a small spatial region, in which it displays
a kinklike minimum. The dark soliton phase distribution however
shows a two-valued distribution with a steep gradient transition
at the kink position. A ''black'' soliton with a stationary zero
density minimum would result for a $\pi $ phase jump, i.e. an
infinite phase gradient.
Starting from the phase distribution difference it becomes clear,
that a dark soliton excitation in a Bose-Einstein condensate can
be produced, by imprinting a phase (typically close to $\pi $) on
one half of the condensate. This was experimentally accomplished
by shining a far detuned laser beam on one half of a trapped
Bose-Einstein condensate. One important aspect in these
experiments was the fact, that the optical resolution of a few
$\mu $m was larger than the condensate healing length, which was
below 500\,nm. This resulted not only in a phase gradient as
compared to a phase jump, but also in a momentum kick for this
position. The momentum introduced into the condensate wavefunction
removed density from the kink position in the form of density
sound waves and thus helped to create the intended dark soliton
excitation. The resulting dynamics is clearly displayed in the
experimental images shown in Sec.~\ref{s:solitons},
Fig.~\ref{f:Darksoliton}. The experiment displays a density and a
dark soliton wave moving in opposite directions and with different
velocities, with the soliton significantly slower than the system
sound velocity.

\section{Physics of ''atom lasers''}~\label{s:atomlaser}
The availability of coherent matter waves with macroscopic
occupation in analogy to the optical laser soon led to the idea of
an ''atom laser'' associated to Bose-Einstein physics. In fact the
phenomenom of Bose-Einstein condensation itself shows many
analogies to laser physics~\cite{Ketterle1997a,Ketterle1998a}. The
modes of the harmonic trapping potential holding the atoms can be
associated with the electromagnetic modes of the light field, with
the trap ground mode being analogous to the laser mode. The atoms
in the higher trap modes can be regarded as gain medium reservoir,
from which they can be transferred to the ground state mode via
stimulated elastic scattering processes in analogy to stimulated
emission from excited atomic states in an optical laser.
Furthermore the optical pumping process could be associated with
the cooling of the sample, leading to an ''oversaturated''
population of excited trap states, similar to population
inversion. These analogies find support in the experimental
observation of condensate wavefunction growth after a sudden
evaporation step in an ensemble with parameters close to the
Bose-Einstein phase transition~\cite{Miesner1998b,Kohl2002a}.
These experiments clearly showed an initial exponential increase
of the condensate wavefunction population, which is consistent
with stimulated scattering processes into this ground state.
Despite the large similarities between Bose-Einstein condensation
and the optical laser, there are also significant differences. One
important aspect is that a Bose-Einstein condensate is a state
reached in thermodynamic equilibrium, while the optical laser
requires working conditions far from this equilibrium. Furthermore
a laser is typically associated with a beamlike output and one
might also wish for a cw operation.

These aspects and according experiments will be discussed in more
detail in the following sections.

\subsection{Ideas and concepts of atom lasers}
The ideas and concepts for atom lasers can be roughly split in two
categories, addressing output coupling and continous loading
processes. One basic idea is to use a trapped Bose-Einstein
condensate as reservoir and to couple part of this matter wave
field out into untrapped states. This is conceptually similar to
partial transmission of the photon field inside a laser resonator
through the outcoupling mirror. Corresponding experiments will be
discussed in the following section. There have been many
theoretical investigations on the physics of the output mechanism
and the resulting matter wave field outside the
trap~\cite{Olshanii1995a,Holland1996c,Wiseman1996a,Ketterle1997a,Wiseman1997a,Dodd1997b,Moy1997a,Moy1997b,Moore1997a,
Steck1998a,Kneer1998a,Band1999a,Breuer1999a,Trippenbach2000a,Bhongale2000a,Jeffers2000a,Wu2000b,Drummond2001a,
Wiseman2002a,Gerbier2001a,Borca2003a,Bradley2003a,Ruostekoski2003a}.
The internal degrees of freedom and their coherent manipulation
with external fields opens more possibilities for this process
than the optical analogue of tunneling through a potential
barrier. The most commonly used mechanism starts from a
magnetically trapped Bose-Einstein condensate, which is
transferred to magnetically untrapped states. Other suggested
schemes reach up to the controlled formation of molecules from an
atomic Bose-Einstein condensate, resulting in a molecular matter
wave laser beam (e.g.~\cite{Borca2003a}).

Other important concepts  aim at an increased matter wave flux by
the realisation of a continously loaded Bose-Einstein condensate.
These ideas ideally complement the above output coupling schemes,
which can only provide a pulsed beam, if starting from a finite
condensate. Several approaches build on the well established
techniques of laser cooling. They try to circumvent typical
temperature and density limitations due to photon scattering and
reabsorption
processes~\cite{Castin1998b,Santos1999a,Santos1999b,Santos2000b,Lewenstein2000a,Bhongale2000a,Floegel2001a,Santos2001a,Olshanii2002a}.
In recent experiments phase space densities on the order of 1/10th
were achieved using narrow linewidth transitions in strontium as
well as by using advanced Raman cooling schemes in optical
lattices~\cite{Vuletic1998a,Kerman2000a,Wolf2000a,Wolf2000b,Treutlein2001a}.

An again different approach relies on continous evaporative
cooling in a continously loaded long magnetic guiding
structure~\cite{Mandonnet2000a}. First experiments on this scheme
demonstrated the pricipal ability to reach dense cold atomic
beams~\cite{Cren2002a} and are very promising to achieve quantum
degeneracy in a beam setup.

Conceptually similar attempts to realise continous or
quasi-continous Bose-Einstein condensation are based on repetitive
or even continous loading of a conservative trap with thermal
atoms and evaporation therein.

The first temporally maintained Bose-Einstein condensate was
however realised by repetitively loading an optical dipole trap
with conventionally produced Bose-Einstein
condensates~\cite{Chikkatur2002a}. Bose-Einstein condensates were
produced with well established standard techniques in a
''production chamber'' and then transferred to a spatially
seperated ''experiment chamber'' with an optical
tweezer~\cite{Gustavson2001a}. In the experiment chamber they were
joined with a continously stored Bose-Einstein condensate. This
experiment demonstrated the feasibility of joining coherent matter
waves for long term observation.

In summary the highly intriguing prospect of a continous ''atom
laser'' has triggered experimental efforts aiming at diverse
schemes and aspects. Until today the goal of a high flux coherent
matter wave source and the interst in deeper understanding of the
properties of atom lasers motivate this work. It should be however
noted, that the current state of the art technology for the
production of Bose-Einstein condensates poses severe limitations
on the practical realisation of orders of magnitude more flux by
simple power arguments. Nowadays the production of coherent matter
waves with a population of $10^7$ atoms every 10\,s requires
typical laser powers of at least 100\,mW for trapping and
precooling a large enough number of atoms. Following a simple
extrapolation even a modest coherent matter wave flux of $10^{10}$
atoms/s (compared to optical lasers with on the order of $10^{20}$
photons/s) would require kW laser power. Future sources will have
to significantly increase efficiencies or realise new
technological approaches to coherent matter waves, e.g. cryogenic
techniques~\cite{Fried1998a}. Once a suitable source is realised
the success of the ''atom laser'' will be determined by our
ability to manipulate and use its properties, which will be
discussed in the following.

\subsection{Experiments on atom lasers}
\begin{figure}
\begin{center}
\includegraphics[width=4cm]{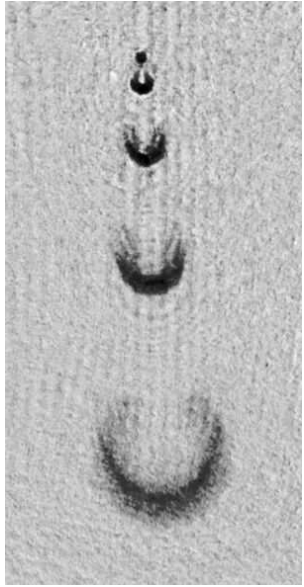}
\end{center}
     \caption{Atom laser pulses realised by repeated pulsed output coupling from
     trapped Bose-Einstein condensate. Image with courtesy of W. Ketterle, MIT.}
     \label{f:Atomlaser}
\end{figure}

The first ''atom laser'' was realised by pulsed output coupling
from a magnetic trapping potential~\cite{Mewes1997a}. In this
experiment a conventionally produced $^{23}$Na Bose-Einstein
condensate in a magnetic trap was subjected to rapid
radiofrequency sweeps. The radiofrequency partially coupled the
condensate wavefunction from the trapped $F=1,\! m_F=-1$ state to
the untrapped $F=1,\! m_F=0$ and antitrapped $F=1,\! m_F=1$
states, resulting in matter wave pulses (see
Fig.~\ref{f:Atomlaser}.

\begin{figure}
\begin{center}
\includegraphics[width=12cm]{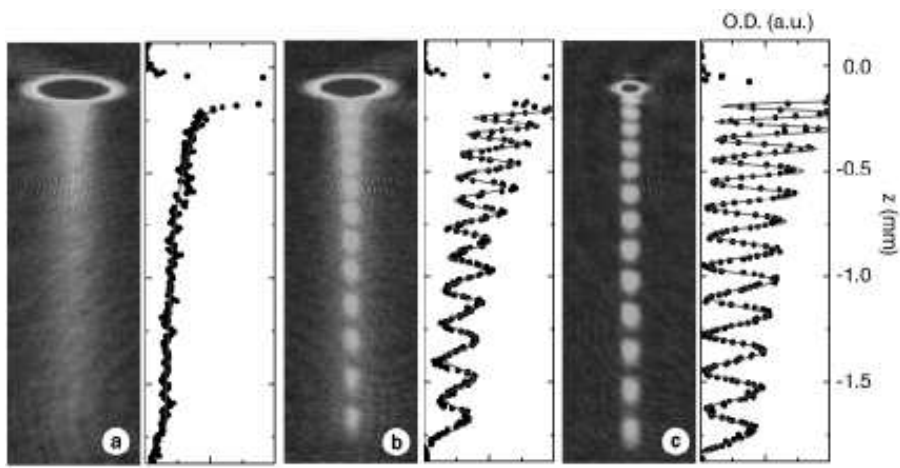}
\end{center}
     \caption{Continous output coupling of two overlapping atomic beams
     from different positions of trapped atomic ensembles. The absorption images
     and the according profiles correspond to a temperature $T>T_c$, $T\leq T_c$ and
     $T\ll T_c$ (from left to right). The buildup of a coherent matter wavefunction
     during the Bose-Einstein phase transition is clearly reflected in the visibility
     of the interference between the two atomic beams. Images with courtesy of
     I. Bloch, Mainz university and T. W. H\"ansch, MPQ.}
     \label{f:Transversecoherence}
\end{figure}

The swept radiofrequency output coupling has the advantage of
reproducible results, even in the presence of a slight magnetic
field fluctuations, but is intrinsically limited to short output
pulses. A significant improvement of the atom laser pulse length
was achieved based on a magnetic trap setup optimised for a very
quiet magnetic field environment~\cite{Bloch1999a}. This enabled
the slow output coupling of an entire Bose-Einstein condensate
with a fixed radiofrequency, leading to a ''quasi-continous atom
laser'' beam. The simultanous output coupling using two radio
frequencies in this highly stable setup resulted in two
overlapping atom laser beams originating from different positions
of the trapped condensate wavefunction~\cite{Bloch2000a}. As these
positions are given by the trap magnetic field variations and the
resonance criteria for rf-induced spinflip transitions, it is
possible to test the coherence of different parts of the
condensate with this techique. The resulting interference pattern
in the according experiment beautifully confirmed the large
transverse coherence length of a trapped Bose-Einstein condensate.
Fig.~\ref{f:Transversecoherence} shows experimental images on the
transverse condensate coherence as a function of temperature.
Further experiments clearly demonstrated, that radiofrequency
output coupling leads to temporally coherent atom laser beams,
which can be used in interferometric
applications~\cite{Kohl2001a}.

A different type of quasi-continous atom laser was demonstrated
based on rapidly pulsed raman output coupling~\cite{Hagley1999a}.
In this experiment a two photon transition was used to partially
transfer the trapped atoms to a different internal and momentum
state. The technique is in principle similar to the Bragg
diffraction scheme discussed above, but with a two photon
transition coupling different internal states. The photon momentum
transfer in this scheme allows the extraction of a beam, aiming in
any direction at will, from the condensate wavefunction, as
opposed to the freely falling beams in the radiofrequency schemes.
Using a rapid succession of short raman pulses resulted in
spatially overlapping atom laser pulses, which joined to one beam.

The analogy between optical and atom laser beams was further
extended by the realisation of a ''mode-locked'' atom
laser~\cite{Anderson1998a}. The realisation of many atom laser
modes was accomplished by first loading a Bose-Einstein condensate
into an one-dimensional optical lattice resulting from a
retroreflected laser beam in the vertical direction. The resulting
condensate was extended over several lattice sites, which were
weakly coupled by tunneling through the optical lattice potential
barriers. Output coupling due to tunneling from all these
potential wells at different heights can be thought of as many
overlapping atom laser beams with different frequencies according
to their velocites aquired in the gravitational potential. Similar
to an optical modelocked laser, in which many resonator modes
overlap to form short spatial light pulses, the output of this
type of atom laser is intrinsically pulsed, with a pulse
seperation connected to the optical lattice spacing and the earth
gravitational potential.

\begin{figure}
\begin{center}
\includegraphics[width=12cm]{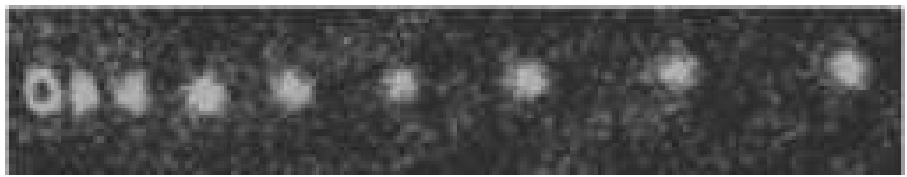}
\end{center}
     \caption{Pulsed atom laser output from a Bose-Einstein condensate (left) stored in
     an optical lattice. (The image is rotated such that gravity acts from left to right.)
     Image with courtesy of M. Kasevich, Stanford university.}
     \label{f:pulsedatomlaser}
\end{figure}

The continous development of new trapping schemes and matter wave
manipulation tools has led to a variety of further pulsed or
quasi-continous atom laser realisations, the most recent of which
stem from optical traps~\cite{Cennini2003a}.

\subsection{Coherence properties of atom lasers}
The idea of an ''atom laser'' as a matter wave source for
experiments, in particular for atom interferometry, has triggered
a vivid discussion about the coherence of this device. This
question depends on the atom laser mechanism under investigation.
For output coupling of matter wave radiation from a trapped
Bose-Einstein condensate there was experimental evidence of
visible interference fringes~\cite{Bloch2000a,Kohl2001a}. This
shows some degree of first order coherence, but the limits of the
coherence length are still subject to dicussion. From the analogy
to the optical laser one would expect the output coupled matter
wave to have a larger coherence length than the size of the
trapped condensate. In this picture it should depend on the output
coupling rate as well as the evaporation rate of the thermal
component, replenishing the condensate. Furthermore there are
influences due to the width of the output coupling transition and
the relative size of the thermal cloud, leading to a phase
diffusion process of the condensate wavefunction.

General aspects concerning the coherence properties of continous
atom laser schemes have so far been investigated for
quasicontinous output coupling~\cite{Kohl2001a}. Particular
coherence aspects connected to continlously pumped atom laser
schemes still lack experimental realisations for their
confirmation.

\subsection{Amplification of coherent matter
waves}\label{s:matterwaveamplification}
 A major point concerning atom lasers and coherent matter wave
sources is the availability of suitable gain media for coherent
amplification. The possibility of matter wave amplification by
stimulated scattering was beautifully revealed by the detailed
investigation of the formation process of a Bose-Einstein
condensate
itself~\cite{Miesner1998b,Gardiner1998b,Bijlsma2000a,Davis2000a,Lee2000a,Kohl2002a}.
This amplification relies on scattering of atoms into the
macroscopically occupied mode of the condensate wavefunction. Due
to the nature of the scattering process particle number, momentum
and energy conservation have to be fulfilled as an essential
requirement for the implementation of a coherent matter wave
amplifier. Evidence of controllable matter wave amplification was
found in a set of spectacular experiments illuminating a
Bose-Einstein condesate with a single laser
beam~\cite{Inouye1999a,Inouye2000a,Schneble2003a}.

In these experiments amplification of spontaneously scattered
photons and the according recoiling matter waves was realised.
Under suitable orientation of light polarisation as well as the
light incident perpendicular to the long axis of a cigar-shaped
Bose-Einstein condensate a regular array of outgoing matter waves
emerges, accomanied with directed light beams containing the
scattered photons. Matter wave amplification using this and
similar schemes is very versatile, not only showing Bragg and
Kapitza-Dirac scattering of atoms in one internal state, but also
Raman scattering with the involved atoms changing their hyperfine
ground state~\cite{Schneble2003a}. These processes clearly
demonstrate the possibility to use a Bose-Einstein condensate
illuminated with one light beam as gain medium for matter wave
seeds, but do not provide a controlled seed, i.e. in comparison
with a light amplifier they would compare to amplified spontaneous
emission (ASE).

\begin{figure}
\begin{center}
\includegraphics[width=8cm]{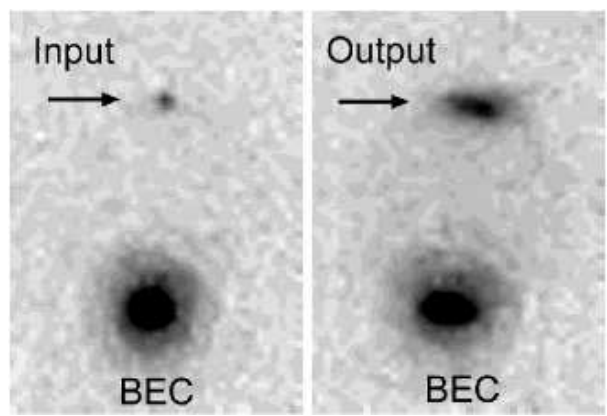}
\end{center}
     \caption{Experimental images on matter wave amplification at MIT.
     The amplification of a small seed matter wave is provided by a
     Bose-Einstein condensate illumiated with an appropriate laser beam.
     The images show the Bose-Einstein condensate and the seeded wave
     some time after the seed has passed the condensate. The left/right image was
     taken without/with the laser beam illumination (no amplification/amplification).
     Images with courtesy of W. Ketterle, MIT.}
     \label{f:Matterwaveamp}
\end{figure}

A full matter wave amplifier based on the above processes was
however realised and characterised by using first a two beam light
pulse to extract a small matter wave seed via Bragg scattering
from a Bose-Einstein condensate and after that apply a single
light beam to turn the same Bose-Einstein condensate into a gain
medium~\cite{Inouye1999b,Kozuma1999b}. In a series of measurements
controlled matter wave amplification using this scheme was clearly
demonstrated with typical amplifications of 10-100 (see
Fig.~\ref{f:Matterwaveamp}). Furthermore the coherence of the
amplification process was proven by applying a second Bragg
scattering pulse after the amplification sequence and
investigating the interference fringes, when the resulting and the
amplified matter wave are overlapped.

In summary there is a well established tool set for matter wave
amplification using Bose-Einstein condensates, its control with
light fields as well as the preparation of seed matter waves
available, which is only limited by the size of the Bose-Einstein
condensate acting as gain medium.

\subsection{Future developments}
The field of coherent matter wave physics building on
Bose-Einstein condensate wavefunctions has seen extraordinary
progress in the last years, with many important developments.
Nowadays there exist standard techniques for the reliable
production of Bose-Einstein condensates as sources of coherent
matter waves and for their manipulation. The intensely pursued
realisation of a continous high flux source for coherent matter
waves will open this dynamic field for interesting applications
and further fundamental experiments. This is in particularly true
for applications in atom interferometry~\cite{Berman1997a}, where
one might expect similar progress as after the introduction of the
laser in light interferometry.

The creation of high precision sensors based on coherent matter
waves is a main motivation for the very active field of atom
chips~\cite{Folman2002a,Dumke2002a}. These devices try to combine
the proven but bulky laboratory techniques for the production of
coherent matter waves with the miniaturisation and control
possibilities offered by the semiconductor industry. A chip based
coherent matter wave sensor would thus greatly improve the
versatility of matter wave experiments for high precision
measurement applications. Recent investigations in this field
concentrate on the coherence properties of 1d waveguides (see
Sec.~\ref{s:phasefluct}) and the influence of fluctuations due to
the presence of the chip surface close to the guided or trapped
matter
waves~\cite{Henkel1999a,Henkel2001a,Kraft2002a,Fortagh2002a,Schroll2003a,Jones2003a,Harber2003a,Lin2004a,Wang2004a}.

In summary the field of atom lasers and their uses is only at the
beginning of a very promising development. Exciting results such
as atom lasers extracted from Bose-Einstein condensates and
coherent matter wave amplifiers as well as the establishment of a
versatile coherent atom optical tool set represent a solid basis
for the future tasks.

\section{Nonlinear atom optics with coherent matter waves}
Nonlinear atom optics adds new fascinating features to the physics
of coherent matter waves. In contrast to light beams coherent
matter waves, even in their realisation as nearly ideal dilute
atomic gases, exhibit intrinsic nonlinear behaviour. The
nonlinearity arises from density dependend interparticle
interactions, e.g. the term $g|\phi(\vec r)|^2$ in the Gross
Pitaevskii equation, Eq. (\ref{GP}). Even if these interactions
only lead to slight changes of the Bose-Einstein phase transition
itself, they add significant new physics to the coherent matter
wave itself. The intrinsic nonlinearity gives rise to many
fascinating phenomena, such as four wave mixing in analogy to
nonlinear optics, solitons in analogy to general nonlinear wave
effects and also new couplings, i.e. leading to spin mixing and
thus magnetic matter wave effects, to name only a few examples. In
the following we will concentrate on these examples and discuss
them in more detail.

\subsection{Four wave mixing}
Nonlinear optics is the basis of intriguing phenomena, such as
wave mixing with sum and difference frequency generation and
became widely accessible with the invention of the laser as
intense and coherent source of light waves. Considering the
discussion on analogies between laser radiation and coherent
matter waves the question arises, if similar phenomena can arise
in these systems. The occurrence of nonlinear phenomena in
coherent matter waves seems natural in view of their intrinsic
nonlineariy: In respect to the comparison with nonlinear optics
the term $g|\phi(\vec r,t)|^2 \phi(\vec r,t)$ describing
interactions in the Gross Pitaevskii equation~\ref{GP} has to be
compared to the term $\chi^{(3)}|E|^2E$ representing the
interaction of the electric field, $E$, in a medium with
susceptibility $\chi^{(3)}$.

\begin{figure}
\includegraphics[width=10cm]{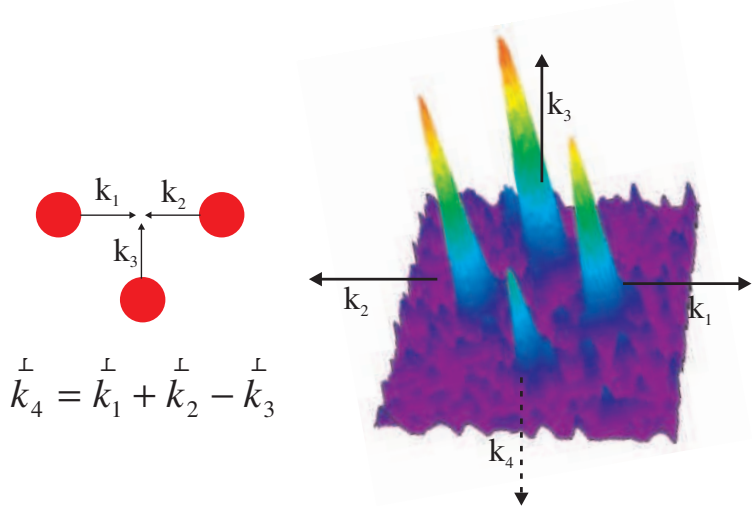}
     \caption{Four wave mixing of matter waves as observed at NIST. Left: schematic
     view of the collision process leading to four wave mixing and the involved
     wavevectors. Right: Experimental absorption image of the matter waves after
     the collision process. Four wave mixing of the three colliding wavepackets gives
     rise to the appearance of a new fourth wavepacket. Experimental image with courtesy of
     the National Institute of Standards and Technology, USA.}
     \label{fig:Fourwavemixing}
\end{figure}

Just as in the discussion on atom lasers and matter wave
amplifiers (see above) there is an important distinction between
matter and light waves in respect to particle number conservation.
While it is possible to convert two photons into a single one with
the sum frequency or to make the reverse transformation in optics,
the same process is not obvious for matter waves due to their
mass. The amount of energy connected to the mass of the atoms
constituting the matter waves discussed in this review is far
beyond all other energy scales involved and thus naturally
enforces particle number conservation. Matter wave mixing
experiments are thus limited to special classes of phenomena, e.g.
degenerate four wave mixing, which was demonstrated in a beautiful
experiment~\cite{Deng1999a} (see Fig.~\ref{fig:Fourwavemixing}).
Transformed to an appropriate frame this experiment realised the
situation. Two counterpropagating matter waves and one orthogonal
matter wave, all of equal momentum, interact with each other and
give rise to a fourth matter wave. This new matter wave has a
wavevector equal to the vectorial sum of the counterpropagating
waves (which is zero) minus the wavevector of the orthogonal wave,
i.e. it is counterpropagating to this wave $\vec k_4 = \vec k_1 +
\vec k_2 - \vec k_3$ (see Fig.~\ref{fig:Fourwavemixing}). Even
though this image is analogous to the case encountered in
nonlinear optics, we want to note again, that this is not true in
general. As an example in the above experimental scheme it would
not be possible to use a different momentum value for the
orthogonal wave, i.e. $k_3 \ne k_1 = k_2$, in order to create
variable output frequencies, as routinely performed in nonlinear
optics. This is a result of energy and momentum conservation for
massive particles and can be most intuitively understood in a
particle scattering based picture for the mixing process.

\begin{figure}
\includegraphics[width=12cm]{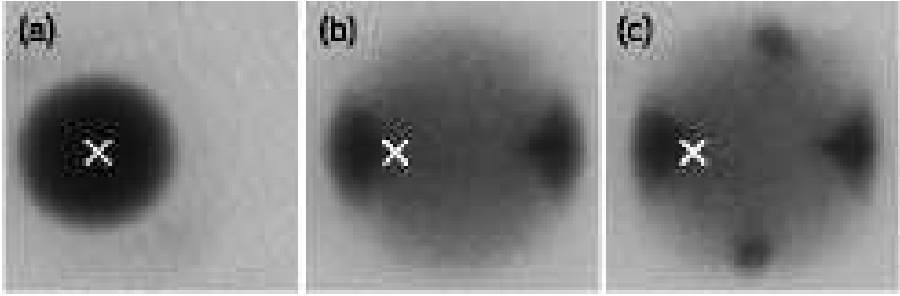}
     \caption{Four wave mixing of matter waves with a small seed and high
     gain. All images display the ensemble after 43\,ms of ballistic expansion.
     (a) The original Bose-Einstein condensate with a 1\%
     seed (barely visible). (b) After the collision of two matter waves created
     by Bragg scattering half the original condensate to a
     higher momentum state. The gray circular background
     corresponds to amplified spontaneous s-wave scattering events
     (similar to ASE in laser physics). (c) High gain four wave
     mixing result after the seed from (a) interacted with the
     colliding waves from (b). Both the seed and the
     counterpropagating four wave mixing outcome were
     significantly amplified and now nearly completely consist of
     correlated atom pairs.
     Image from~\cite{Vogels2002a} with courtesy of W. Ketterle, MIT.}
     \label{fig:swaveStreuung}
\end{figure}

In this approach the counterpropagating matter waves with
wavevectors $\vec k_1$ and $\vec k_2$ interact via elastic
collisions coupling the waves to a s-wave scattering sphere (only
s-wave collisions are involved due to the low relative momenta)
(see Fig.~\ref{fig:swaveStreuung} (b)). In a single particle
picture this sphere results from the quantum mechanical average
over many binary collisions. Each binary collision however has to
fulfill momentum and energy conservation. Momentum conservation
immediately leads to $\vec k_1 + \vec k_2 = \tilde{\vec{k_1}} +
\tilde{\vec{k_2}}$ and due to $\vec k_1 + \vec k_2 =0$ to
$\tilde{\vec{k_1}} = - \tilde{\vec{k_2}}$ as well as $\tilde k_1 =
\tilde k_2$. Energy conservation on the other hand requires $k_1^2
+ k_2^2 = \tilde k_1^2 + \tilde k_2^2$ resulting in
$k_1=k_2=\tilde k_1 = \tilde k_2$. The collisional coupling thus
can only support equal and opposite momenta for the waves $\vec
k_3$ and $\vec k_4$.

At this point it is important to realise, that the collisional
view of the four wave mixing process makes an intriguing
connection to matter wave amplification as discussed in
section~\ref{s:matterwaveamplification}. The emergence of a
clearly visible peak corresponding to $\vec k_{\rm new} = -\vec
k_3$ instead of a scattering sphere can be interpreted as an
amplification process of the orthogonal matter wave due to
stimulated scattering. Momentum conservation then necessariliy
requires the emergence of a fourth matter wave with opposite
momentum (see Fig.~\ref{fig:swaveStreuung}. The amplification
properties of the four wave mixing process were investigated in
detail in~\cite{Vogels2002a}. They also point out, that the
degenerate matter wave scattering leads to entanglement between
the amplified and the new wave components: Each binary scattering
event leads to two particles moving in opposite directions with no
possibility to decide which of the colliding particles took which
way. If the seed wave $\vec k_3$ has relatively small occupation
in comparison to its amplified population, then the new and the
amplified wave might be imagined to essentially consist of
entangled atom pairs.

As a last point we want to stress, that the amplification process
in four wave mixing experiments can be interpreted in an analogous
way to the amplification using Bragg scattering off a light
grating as discussed in section~\ref{s:matterwaveamplification}.
Indeed the matter waves with wavenumbers $k_1$ and $k_3$ build up
a matter wave density grating by which the third matter wave $k_2$
is scattered as in light optics into the new wave $\vec k_4$.

In summary four wave mixing of coherent matter waves represents a
fascinating analogy to wave mixing phenomena in nonlinar optics.
In addition the entanglement creation connected to the four wave
mixing process leads to a new source of entangled particles with
promising applications in quantum computing or high resolution
interferometry~\cite{Bouyer1997a,Vogels2002a}.

\subsection{Dark solitons}\label{s:solitons}
Another class of nonlinear phenomena in matter wave physics
consist in solitons as fundamental excitations. Also with the
physics of solitons matter wave physics is linked to general
nonlinear wave effects, where solitons appear in a wide variety of
systems, e.g. in optical fibers or as ''Tsunamis'' in water. The
word soliton denotes the phenomenon of single, i.e. solitary,
wavepacket like excitations, which travel with constant shape and
without significant dissipation. In more physical terms they are
density variations which self stabilise due to a balance of
nonlinear interactions (trying to conpress the wavepacket) and
dispersive effects (tending to spread the wavepacket) .

In a Bose-Einstein condensate matter wave this balance occurs
between the nonlinear mean field interaction and quantum pressure.
The quantum pressure is related to the kinetic energy associated
with the soliton velocity field, and is minimized by widening the
spatial extend of a density variation. It is important to note,
that this is independend of the nature of the density variation,
i.e. whether it is a denisty maximum or minimum. The possible
solitonic matter wave excitations can thus be classified according
to the sign of the s-wave scattering length, $a$, determining
wether the interparticle interaction is attractive ($a<0$) or
repulsive ($a>0$) leading to bright or dark solitons.

A {\it bright soliton}, i.e. a stable density maximum, can occur
for the case of attractive interactions, which tend to further
increase density by compressing the maximum and thus balance the
spreading effects. Bright solitons and trains of bright solitons
have been observed in experiments with bosonic Lithium, having
attractive interactions~\cite{Khaykovich2002a,Strekker2002a}.
These experiments are very demanding, as large Bose-Einstein
condensates with attractive interactions are in general not stable
against collapse. The experimental realisation was possible by
tuning the scattering length with a Feshbach resonance, first
producing a condensate with repulsive interactions, loading it
into a one-dimensional geometry and then switching the interaction
to attractive by a change in the magnetic offset field.

A {\it dark soliton}, i.e. a stable density minimum, is a
fundamental excitation for matter waves with repulsive
interactions, which are realised in most experiments on
Bose-Einstein condensation. Here the quantum pressure spreading
effect can be balanced by the repulsive interactions, for the case
of a density minimum. The interaction energy would be minimised by
filling this minimum, which is by shrinking its spatial extends.
In the following we will focus on this kind of soliton which was
first experimentally realised at Hannover as well as
NIST~\cite{Burger1999a,Denschlag2000a}, and later also studied at
JILA~\cite{Anderson2001a}.

For a homogeneous, one dimensional Bose-Einstein condensate matter
wave function dark soliton excitations can be found as the
following solution of the 1d Gross Pitaevskii equation
\cite{Zakharov1972a,Fedichev1999b}:
\begin{equation}
\Psi_k(x) = \sqrt{n_0}\left( i\frac{v_k}{c_s}+
\sqrt{1-\frac{v_k^2}{c_s^2}}
\tanh\left[\frac{x-x_k}{l_0}\sqrt{1-\frac{v_k^2}{c_s^2}}\right]\right)
. \label{soliton_eqn}
\end{equation}
In this equation the parameters are the matter wave density $n_0$,
the position $x_k$ and velocity $v_k$ of the dark soliton, the
correlation length $l_0=(4\pi a n_0)^{-1/2}$, and the speed of
sound $c_s=\sqrt{4\pi a n_0}\hbar/m$, where $m$ is the atom mass.
Analysing this expression one finds the special case of a so
called {\it black soliton} with zero density at the minimum, if
the matter wave phase exhibits a $\pi $ phase jump at the density
kink position. The black soliton is at rest, i.e. $v_k=0$, while
for all other cases the kink in the condensate wave function moves
along the condensate, has nonzero density and is associated with a
finite phase gradient.

The finite kink velocity of a dark soliton is connected to the
gradient in the matter phase distribution $\phi(\vec{r}\,)$, which
defines defines a probability current density. The associated
superfluid velocity field
$v(\vec{r}\,)=\frac{\hbar}{m}\nabla\phi(\vec{r}\,)$ leads to a
localized peak in the velocity distribution of the matter wave.

The relatively simple two-valued phase distribution of a dark
soliton makes this excitation favourable for the method of phase
imprinting as discussed in Sec.~\ref{s:phasemanip}. This method
was used for the first experimental realisation of dark soliton
excitations in Bose-Einstein
condensates~\cite{Burger1999a,Denschlag2000a}.

An important aspect is that although dark solitons are only stable
in one dimensional or quasi one dimensional systems, they can also
be produced in three dimensional near spherical
geometries~\cite{Denschlag2000a,Anderson2001a}. In these
geometries they are however unstable against transverse
excitations~\cite{Fedichev1999b,Muryshev1999a}, e.g. their decay
into vortex rings was observed in a recent
experiment~\cite{Anderson2001a}.

\begin{figure}
\begin{center}
\includegraphics[width=12cm]{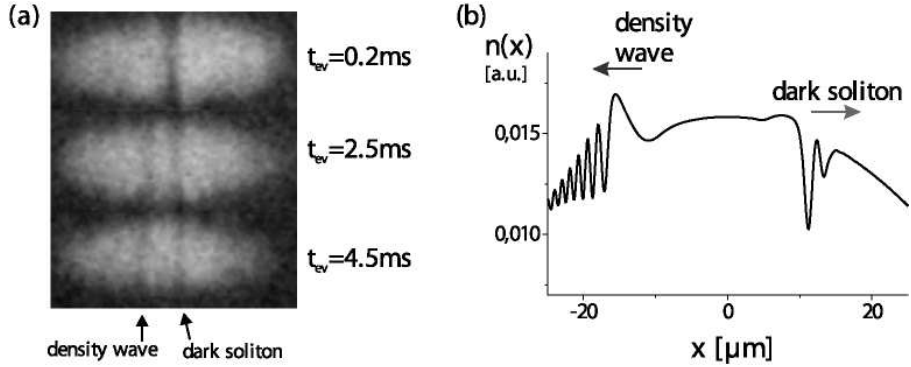}
\end{center}
     \caption{(a) Evolution of a dark soliton in a Bose-Einstein condensate wavefunction.
     The matter wave evolved in the trapping potential for a time, $t_{ev}$ after phase imprinting.
     The images were taken after an additional time-of-flight of 4$\, $ms. The matter wave dynamics
     during this additional time of flight leads to density minima for the dark soliton as well as for
     an additionally created density wave~\cite{Burger1999a}. (b) Numerical simulation of the
     density distribution of a dark soliton propagating in a BEC after phase imprinting.}
     \label{f:Darksoliton}
\end{figure}

Fig.~\ref{f:Darksoliton} shows experimental data and numerical
simulations on the propagation of a dark soliton during some
evolution time, $t_{ev}$, in the trap.

In addition to the dark soliton a density wave is created in the
phase imprinting process due to the finite optical resolution as
compared to the matter wave healing length. The resulting steep
light shift potential gradient transfers momentum onto the part of
the matter wave, which is at the intended kink position. This
reduces the density at the position of the imprinted phase
gradient and thus helps in the creation of the kink associated
with the dark soliton.

The velocity field due to the phase gradient of the dark soliton
is directed opposite to the movement of the density kink. This can
be intuitively understood as the movement of the density minimum
necessitates a transfer of matter in the opposite direction.

The matter wave flux underlying a dark soliton was analysed using
Bragg velocity spectroscopy, i.e. Bragg scattering in the regime
of long pulses as discussed in Sec.\ref{s:minimumuncertainty}. In
these experiments good agreement with theory was
found~\cite{Bongs2003a}.

The creation of solitons in Bose-Einstein condensates demonstrated
fundamental nonlinear excitations in matter waves and represents a
nice experimental implementation of the method of phase
imprinting. The investigation of solitons as signature of
nonlinear excitations depending on the interparticle interactions
is an actual topic in the physics of matter waves. Recent studies
demonstrated the possibility use optical lattice geometries in
order to modify the effective mass and thus the interactions of
the matter waves~\cite{Eiermann2003a}. With this dispersion
management technique it was possible to change the nature of
interactions from repulsive to attractive and to observe bright
solitons in this regime. Further progress in this area is expected
from the realisation of so called discrete solitons in optical
lattices, which are an actual topic of theoretical
work~\cite{Trombettoni2001a,Abdullaev2003a,Louis2003a,Ostrovskaya2003a,Ahufinger2003a}
as well as mixed species solitons.

\subsection{Magnetic matter wave effects}
The intrinsic nonlinear interactions of matter waves not only
gives rise to the fascinating nonlinear single state effects
discussed above, but opens a wide variety of new phenomena based
on an inter-state coupling in multi-component condensates.
Multi-component matter waves are composed of atoms with multiple
internal states, e.g. multiple hyperfine states or multiple
magnetic quantum states in atoms with non-zero total angular
momentum.

In the typical magnetic trap arrangement for the creation of
Bose-Einstein condensates the additional degrees of freedom due to
the internal atomic states are frozen by energetic restrictions.
Once released by hf coupling~\cite{Matthews1998a} or by storage in
a state-independent optical dipole trap~\cite{Stamper-Kurn1998b}
magnetic effects and intriguing multi-state dynamics as well as
new thermalisation phenomena become visible.

\begin{figure}
\begin{center}
\includegraphics[width=12cm]{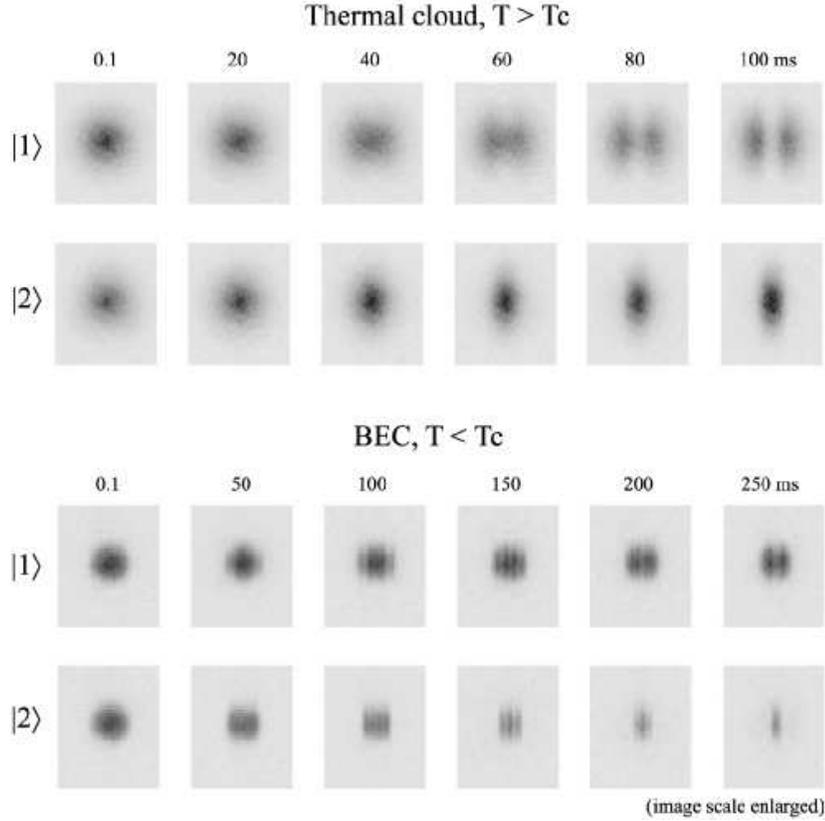}
\end{center}
     \caption{Dynamics of the component separation and domain formation in a
     quasi spin-1/2 spinor gas (states $|1\rangle $ and $|2\rangle $,
     for the case of a thermal gas ($T>T_c$) and a nearly pure Bose-Einstein
     condensate ($T<T_c$). This example shows the different timescales for spin
     transport for the two regimes, which are again significantly modified in a
     Bose-Einstein condensate with significant normal component~\cite{McGuirk2003a}.
     Image with courtesy of E.~A. Cornell, JILA, Boulder}
     \label{f:spinseparation}
\end{figure}

Experiments on coherent multi-component matter waves with spin
degree of freedom were performed in two different systems. So
called spinor condensates were first realised with optically
trapped $^{23}$Na in the F=1 hyperfine
state~\cite{Stamper-Kurn1998b,Miesner1999a,Stamper-Kurn2001a,Leanhardt2003b},
while at the same time a different class of multi-component system
was realised by coupling the magnetically trapped $F=1, m_F=-1$
and $F=2, m_F=1$ hyperfine states of
$^{87}$Rb~\cite{Matthews1998a}. This system is analogous to an
effective spin 1/2 system and shows fascinating mixing and
demixing effects~\cite{Hall1998a}. Long intrinsic inter-state
coherence times up to several seconds were
found~\cite{Hall1998b,Matthews1999a}. In recent experiments this
system was used for various studies on spin transport, coherence
and decoherence phenomena, in particular with respect to
interactions with the normal
component~\cite{McGuirk2002a,Harber2002a,McGuirk2003a,Lewandowski2003a}.
Fig.~\ref{f:spinseparation} shows an example of spin domain
formation, which was studied in detail in~\cite{McGuirk2003a}.

Recently studies on $F=2$ $^{87}$Rb spinor condensates were
performed~\cite{Schmaljohann2004a} that extend the multi-component
physics to five states and offer additional rich structures.

Optically trapped spinor condensates offer the possibility of
collisional coupling, which gives rise to intrinsic magnetic
properties of the system and opens intriguing new possibilites for
the study of finite temperature phenomena.

The magnetic properties of coherent matter waves are typically
considered for a single hyperfine state, e.g. denoted by the total
single particle spin $F$. They give rise to a characteristic
distribution of the $m_F$-components, i.e. the spin orientations
projected on a quantisation axis, and can be classified similar to
solid state magnetic effects.

Although studies on dipolar quantum gases, where magnetic (or
electric) dipole-dipole interactions play a dominant role, are an
actively pursued
field~\cite{Goral2000a,Santos2000a,Giovanazzi2002a,Damski2003a,Yi2003a},
most experiments deal with matter wave systems dominated by
collisional mean field interactions. In the following we will
concentrate on these systems, ruled by a spin-dependence of the
molecular potential curves which are involved in the collisional
interaction. The interactions can be classified most intuitively
by first considering the collision of two particles, each with
total spin $F$~\cite{Ho1998a,Koashi2000a,Ueda2002a}. The
two-particle collision now only depends on the total two-particle
spin $f$, which can take values from $0..2F$ depending on the
orientation of the single particle spins. Note that only even
values of $f$ occur due to bose symmetry. The resulting two
particle interaction potential can thus in principle be divided
into the corresponding contributions by the introduction of the
s-wave scattering lengths $a_f$ for the different two-particle
spin collisions~\cite{Ho1998a,Koashi2000a}:
\begin{equation}
\hat V(\mathbf{r_1-r_2})=\delta(\mathbf{
r_1-r_2})\sum_{f=0}^{2F}\frac{4\pi \hbar^2a_f}{m}\hat P_f .
\end{equation}
Here $\hat P_f$ is the projection operator onto total spin $f$ and
$m$ is the mass of a single atom. A further simplification
concerning the experimentally relevant case of a coherent matter
wave with particle spin $F=1,2$ finally results in the following
spin-dependend mean field energy contributions:
\begin{equation}
K_{spin}=c_1\langle \mbox{\boldmath$F$} \rangle^2+\frac{4}{5}
c_2|\langle s_- \rangle|^2-\tilde p \langle F_z\rangle-q\langle
F_z^2\rangle .
\end{equation}
In this expression $\langle \mbox{\boldmath$F$} \rangle $,
$\langle F_z\rangle $ and $\langle s_- \rangle $ denote the
expectation values for the single particle spin vector, its $z$
component and the spin-singlet pair amplitude.

The parameters $c_1$ and $c_2$ characterise the magnetic
properties of the system and are essentially given by linear
combinations of the spin-dependend s-wave scattering lengths. For
F=1 systems $c_1=\frac{4\pi\hbar^2}{m}\times\frac{a_2-a_0}{3}n$
and $c_2=0$, while for F=2 systems
$c_1=\frac{4\pi\hbar^2}{m}\times\frac{a_4-a_2}{7}n$ and
$c_2=\frac{4\pi\hbar^2}{m}\times\frac{7a_0-10a_2+3a_4}{7}n$, with
the particle density $n$.

Considering only these parameters the spinor systems can by
classified into antiferromagnetic (or polar), ferromagnetic and
cyclic behaviour. The first two phases occur for both $F=1$ and
$F=2$ systems, while the cyclic phase needs the additional degrees
of freedom only offered in the $F=2$ system. An antiferromagnetic
or polar system shows miscibility of the $m_F=-F$ and $m_F=+F$
spinor components, while they spatially seperate for a
ferromagnetic system. The cyclic phase supports a mixture of the
$m_F=-2$, $m_F=0$ and $m_F=+2$ spins~\cite{Ueda2002a}. The
corresponding ground state matter wave distributions additionally
depend on experimental parameters, in particular the magnetic
offset field and the magnetic field gradient. Detailed phase
diagrams taking these effects into account can be found for $F=1$
systems in~\cite{Stamper-Kurn2001a}.

One important point we want to emphasize for atomic systems
dominated by the above interactions (neglegible dipole-dipole
interactions) is that the total spin projection on the
quantisation axis is conserved in a two-particle scattering event.
This introduces important differences for spin dynamics in
coherent matter waves as compared to condensed matter systems. In
condensed matter the spin degree of freedom is coupled to other
degrees of freedom of the bulk system, which can cause spin flip
processes and typically limits coherent dynamics.

In the spinor coherent matter waves discussed above spin dynamics
is fully due to their intrinsic quantum mechanical interactions.
Consequently spin dynamics in $F=1$ spinor gases is determined by
the process $|-1\rangle + |+1\rangle \leftrightarrow |0\rangle +
|0\rangle$, (denoting the states by their $m_F$-component for
simplicity). Although similar rules apply for $F=2$ systems, the
availability of 5 spin components allows much more complex
dynamics in this case.

A fascinating set of experiments investigated the spinor
properties and dynamics of the $F=1$ state of
$^{23}$Na~\cite{Stenger1999a,Stamper-Kurn2001a}. These experiments
showed that this state behaves antiferromagnetic and studied the
corresponding phase diagram. Miscible and immiscible spin mixtures
were produced and tunneling processes during their spatial
rearrangement were studied~\cite{Stamper-Kurn1999b}. Furthermore
metastability and delay effects were observed in spinor
dynamics~\cite{Miesner1999a}. Recent experiments showed, that the
$^{23}$Na system shows severe limitations for studies of the $F=2$
spinor system due to rapid hyperfine loss
processes~\cite{Gorlitz2003a}.

\begin{figure}
\begin{center}
\includegraphics[width=12cm]{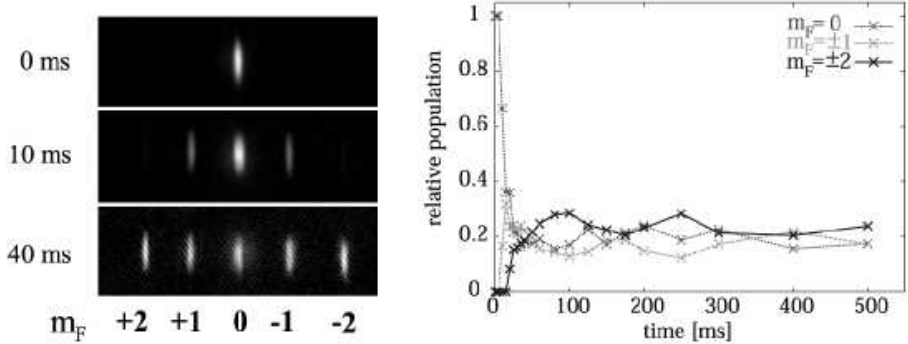}
\end{center}
     \caption{Example for F=2 spin dynamics starting with a matter wave
     prepared in the $m_F=0$ state. The absoption images on the left show
     the temporal evolution of the individual spin components, which overlap
     in the trap but are seperated by a Stern-Gerlach setup for the images.
     First the $\pm 1$ components are populated and only later also the $\pm 2$
     spin components. This behaviour is also reflected in the graph on the right,
     which furthermore shows oscillatory behaviour in the magnetic dynamics of the
     quantum gas.}
     \label{f:spindynamics}
\end{figure}

The first studies on a $F=2$ spinor system were achieved in
$^{87}$Rb (see Fig~\ref{f:spindynamics}) showing intriguing spin
dynamics with evidence for coherent oscillations, condensate
melting thermalisation effects and an antiferromagnetic ground
state behaviour, close to the cyclic phase with metastable states
belonging to this phase~\cite{Schmaljohann2004a}. Furthermore it
was shown that the $^{87}$ $F=1$ state behaves
ferromagnetic~\cite{Chang2003a,Schmaljohann2004a}.

In conclusion the nonlinear interstate coupling in matter waves
gives rise to new fascinating phenomena and magnetic properties.
Recent experiments provide a variety of rich spinor systems, with
exciting physics to be explored. These systems offer promising
extension possibilities by new matter wave species (e.g. $F=3$ and
$F=4$ in $^{85}$Rb and $^{133}$Cs) but also by coherent coupling
of different hyperfine states or tuning with recently observed
interstate Feshbach resonances~\cite{Erhard2003a}.

\section{Guiding structures and optical elements for coherent
matter waves}\label{s:guiding} One main point of interest in
coherent matter waves lies in their application, in particular in
high precision sensors for fundamental physics as well as
commercial tasks. These prospects are strongly supported by the
analogies with optical laser beams but the mass and internal
atomic structure lead to fundamental differences. These
differences open new prospects for applications, but also require
adapted strategies for matter wave manipulation. In terms of
counteracting gravitational acceleration and prolonging
observation times waveguides for matter waves seem highly
promising. In this respect it is desirable to achieve guiding
structures with high transverse confinement, as this enhances
shielding from external noise and facilitates single transverse
mode guiding. In addition the coherent manipulation of matter
waves is a nontrivial task and requires considerable
investigation.

\subsection{Magnetic guides}
Magnetic guiding structures for matter waves have been under
research from the beginning of atom optics~\cite{Folman2002a}.
They rely on the same principles (i.e. the Zeeman energy shift of
suitable atomic sublevels) as magnetic traps commonly used in
Bose-Einstein condensation experiments. Magnetic matter waveguides
have been demonstrated with different geometries, e.g. in
wire-lined hollow fibers~\cite{Key2000a} and along current
carrying
wires~\cite{Schmiedmayer1995a,Fortagh1998a,Denschlag1999a}. The
latter realisation is particularly well suited for miniaturisation
and commonly used in various atom chip designs. It relies on the
creation of a two-dimensional magnetic quadrupole field by
combining the magnetic field of a current carrying wire with a
homogeneous offset field. The scaling of the field gradient with
the wire dimensions and the distance from the wire results in
stronger possible confinement for miniaturised setups as an
advantage for atom chip designs. A variety of elongated traps and
guiding structures have been realised with this
design~\cite{Weinstein1995a,Reichel1999a,Muller1999a,Dekker2000a,Folman2000a,
Drndic2001a,Ott2001a,Hansel2001a,Leanhardt2002a,Schneider2003a}.
Transverse confinements with an energy spacing between ground and
first excited state up the h times 100\,kHz range can nowadays be
achieved and trapping frequencies up to the MHz range seem
feasible in the near future~\cite{Reichel2003a}.

The proximity of a room temperature surface in the miniaturised
atom chip setups introduces new heating effects, which were
proposed by~\cite{Henkel1999a,Henkel2001a,Schroll2003a,Wang2004a}
and also recently investigated in several
experiments~\cite{Jones2003a,Harber2003a,Lin2004a}. The main
effects arise from magnetic fields due to thermal current noise in
metallic substrates and evanescent electromagnetic waves above
dielectric substrates. The latter effect is much smaller and does
not seem to limit current technology but metallic surfaces were
found to induce significant heating and trap loss for samples at
close distance.

Further influences on guided or trapped atomic samples were found
due to the current carrying wire
itself~\cite{Kraft2002a,Fortagh2002a,Leanhardt2002a}. These
effects are found to be connected to the current flow, which might
be made irregular by small fabrication irregularities and/or
crystal grain boundaries inside the conductor.

Wire based guides and atom chips are very versatile devices with
many promising applications. The achievement of significant
further miniaturisation is an acitvely pursued technological task,
which has to take into account fundamental surface induced heating
effects as well as find suitable materials and processes to ensure
regular current flow in the guiding wire.

\subsection{Optical guides}
Optical guides represend a second promising class of waveguides
for coherent matter waves. They are based on the dipole potential,
i.e. the ac Stark shift of the atomic states, induced by the
interaction of the atoms with laser light with a frequency far
detuned from the atomic resonance (see e.g. ~\cite{Grimm2000a}).
In a simple model the dipole potential is given
by~\cite{Grimm2000a}:
\begin{equation}
  U_{\rm dip}(\vec r) = -\frac{3\pi c^2}{2\omega_0^3}\left(
  \frac{\Gamma }{\omega_0-\omega } +\frac{\Gamma }{\omega_0+\omega
  }\right)^2 I(\vec r).
\end{equation}
In this equation $\omega_0 /2\pi$ is the atomic resonance
frequency, $\omega /2\pi $ is the laser frequency and $\Gamma $ is
the damping rate corresponding to the spontaneous decay rate of
the upper level. typical depth of dipole potential traps and
guides correspond to $\mu $K to mK, if expressed as temperatures.

Light can guide atoms in a very flexible way and a high
confinement not necessarily relies on proximity to surfaces
possibly leading to heating effects.

A straightforward method for the realisation of an optical matter
wave guide is to rely on the potential created by a freely
propagating laser beam. The detuning of the laser frequency versus
the atomic resonances determines, wether the atoms are confined at
positions of high (for red detuning, i.e. laser frequency below
the atomic resonance frequency) or low (for blue detuning, i.e.
laser frequency above the atomic resonance frequency) light
intensity.

The case of red detuning was mainly used for the realisation of
matter wave
traps~\cite{Askaryan1962a,Letokhov1968a,Chu1986a,Stamper-Kurn1998b}
in the focus of a laser beam. The trapping effect instead of a
pure guide is due to the propagation rules for the evolution of a
gaussian laser beam. These rules lead to a stronger divergence of
the beam the smaller its focus is. In other words creating a
strong transverse confinement by tightly focussing the trapping
beam, will lead to an additional axial intensity gradient, which
provides a confinement in beam direction. This effect can be
minimised by the use of weakly focussed laser beams of high power,
but is limited by the available lasers.

\begin{figure}
\begin{center}
\includegraphics[width=6cm]{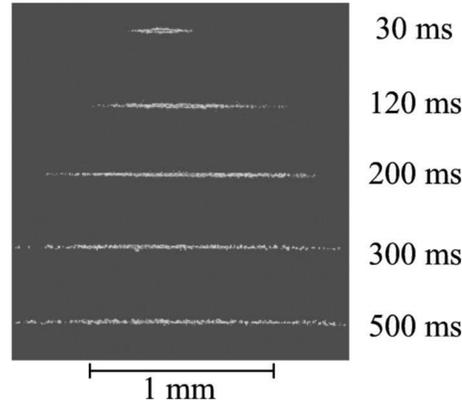}
\end{center}
     \caption{Expansion of a Bose-Einstein condensate inside a doughnut-shaped,
     far blue detuned light field acting as a matter waveguide~\cite{Bongs2001a}.}
     \label{f:donut}
\end{figure}

Optical waveguides based on a hollow blue detuned laser
beam~\cite{Noh2002a,Noh2002b}, creating a repulsive potential tube
for were demonstrated for thermal
atoms~\cite{Gallatin1991a,McClelland1991a,Kuppens1996a,Kuga1997a,Kuppens1998a,Xu1999a,Song1999a,Xu2001a}
and coherent matter waves~\cite{Bongs2001a}. In~\cite{Bongs2001a}
a TEM$_{00}$ mode laser beam was transformed into a
doughnut-shaped hollow TEM$_{01}^*$ laser beam using a
blazed-grating phase hologram. The beam parameters were chosen as
to lead to a Rayleigh range on the order of 1\,mm, which has to be
compared to the initial 100$\, \mu $m size of the Bose-Einstein
condensate wavefunction loaded into this guide. This resulted in a
slight longitudinal potential hill in the guide as opposed to the
potential minimum in red detuned beams. Fig.~\ref{f:donut} shows
the slow spreading and propagation of a matter wave in this
guiding geometry. It was shown, that this spreading is consistent
with a transformation of mean field energy into one dimensional
kinetic energy along the guide direction. Furthermore Bragg
interferometric measurements demonstrated the coherence of the
sample after some expansion time inside the guide.

Note that in contrast to the above low temperature coherence
measurements, phase fluctuations occur at finite temperatures in
these elongated geometries and pose severe limitations on
applications (see Sec.~\ref{s:phasefluct}). Successful guided
measurement applications seem to require operation at temperatures
much below typical Bose-Einstein phase transition temperature.

Other actively pursued methods for the realisation of optical
waveguides concentrate on integration and flexible geometries. One
important scheme relies on the guidance of atoms in light fields
themselves guided in hollow
fibers~\cite{Renn1995a,Renn1996a,Ito1996a,Renn1997a,Ito1997a,Yin1998a,Muller2000a,Hayashi2003a}.
This promises to remove the contraints of gaussian optics for
freely propagating beams and to realise long, curved guiding
structures with spatially constant and high confinement.
Successful guiding was demonstrated by confining atoms in the
evanescent field of blue detuned light modes guided in the glass
fiber tube. Technical limits of this technique lie in speckle
patterns inside the fiber due to multimode operation, which lead
to heating and atom loss. Furthermore suitable loading schemes
funneling the atoms into the repulsive laser field have to be
applied. A more natural loading seems possible with red detuned
guiding light fields, but these experience severe losses inside
conventional hollow fibers. The recent invention of hollow
photonic crystal fibers might change this situation in future
experiments.

New approaches also combine atom chip technology with optical atom
guiding~\cite{Birkl2001a}. Recent experiments realised trapping
and guiding of atoms in light fields shaped with microlens arrays
and cylindrical microlenses~\cite{Dumke2002a,Dumke2002b}. These
experiments also demonstrated integrated beamsplitters and
interferometric setups.

Optical atom guiding in its various forms shows high flexibility
and comparable confinement strength as its magnetic counterpart.
Ongoing technological developments, in particular microoptics and
photonic crystal materials offer promising prospects for reliable
coherent matter waveguides.

\subsection{Mirrors and beamsplitters}
The ability to manipulate and control coherent matter waves relies
on the availability of suitable optical
elements~\cite{Adams1994a}. In comparison to light optics, the
situation in matter wave optics is complicated by several aspects.
Some intrinsic points are the different dispersion relation for
matter waves, which often leads to velocity dependent,
i.e.~dispersive processes. Furthermore the interparticle
interaction causes nonlinear processes to influence matter wave
dynamics at high densities.

In addition to the guiding structures for matter waves mirrors and
beamsplitters as basic optical elements are required for the
implementation of applications such as interferometric
measurements. These elements have to preserve the coherence of the
matter waves, i.e. to avoid spontaneous processes in light based
realisations. In principle all atom optical elements are either
based on suitable potential geometries, e.g. short range repulsive
potentials for the realisation of mirrors, or on coherent atom
light interations changing the momentum and/or the internal state
of the undelying atoms. The potentials are typically realised with
light fields or magnetic fields.

The reflection of atoms with relatively high kinetic energy has
been demonstrated using steep potentials realised with evanescent
waves~\cite{Balykin1988a,Kasevich1990a,Aminoff1993a,Szriftgiser1996a,Landragin1996a,Savalli2002a}
and periodic magnetic surfaces
\cite{Roach1995a,Sidorov1996a,Johnson1998a,Lau1999a,Drndic1999a,Lison1999a,Saba1999a,Hinds1999a,Rosenbusch2000a,Bertram2001a,Arnold2002a}.
These elements however critically depend on their surface
roughness, which often hinders a purely specular reflection.

These aspects can be relatively well controlled by the reflection
of coherent matter waves from flat light fields~\cite{Bongs1999a}
but in this case the gradient of the light field potential is
weaker leading to a ``softer'' mirror compared to evanescent
waves. As an advantage these mirrors do not require the presence
of a surface, i.e. they can also transmit matter waves.

A different type of ''soft'' mirror has been demonstrated for
Bose-Einstein condensates and atom-lasers with magetic
fields~\cite{Bloch2001a}.

Very controlled and non-dispersive mirrors as well as variable
beamsplitters have been realised using Bragg reflection in the
short pulse regime (see Sec.~\ref{s:minimumuncertainty} ans
Sec.~\ref{s:Bragginterferometer}) as well as Raman
reflection~\cite{Berman1997a}. Raman reflection is similar to the
Bragg process discussed above, but in this case the internal
atomic state is changed during the atom-laser interaction. Due to
their high reliability and non-dispersive character these matter
wave optical elements are used in most coherent matter wave
interferometric schemes which will be discussed in the next
section. Their main drawback lies in the small possible momentum
transfer, limited to a few photon recoil momenta.

Guiding structures open a new way to realise a beamsplitter by
splitting a matter wave guide in two new
branches~\cite{Cassettari2000a,Dumke2002b}. The question on the
coherence properties of this new type of beamsplitter and their
possible use in interferometric structures is currently under
investigation~\cite{Andersson2002a,Kreutzmann2003b}

The search for coherent high momentum transfer non-dispersive
matter wave optical elements is still an active field of research,
which has strong impact on future applications.

\section{Atom interferometry with coherent matter
waves}\label{s:atominterferometry} There are many expectations
connected to the introduction of coherent matter waves into high
precision interferometric measurements. These expectations are
supported by the analogy to light optics and the rapid gain in
sensitivity after the invention of the laser as source of coherent
light waves. The interest in atom interferometric sensors stems
from the high sensitivity which can be realised with these
instruments. State-of-the-art atom interferometers are among the
most precise sensors for the measurement of frequency, rotations,
gravity and gravity
gradients~\cite{Berman1997a,Ruschewitz1998a,Gustavson2000a,Peters2001a,Becker2001a,Udem2001a,Wilpers2002a,McGuirk2002b,Takamoto2003a,Kolachevsky2004a,Roberts2004a}
even with a thermal atom source. The prospect of significant
further sensitivity increases due to the implementation of a
coherent matter wave source is connected to several issues, which
are under active investigation.

First of all the progress in light interferometry with the
invention of the laser is to a great part due to an increase in
photon flux in the interfering mode. Modern atom interferometric
schemes are able to extract a signal from entire thermal ensembles
of atoms, given that their momentum and spatial spread is not too
large. In this respect a coherent matter wave source has to
compete with thermal atom sources which nowadays tpically have
many orders of magnitude higher flux. Thermal atom beam sources
achieve a typical flux of $10^{14}$ atoms/s and laser cooled
sources achieve in the range of $10^9$ atoms/s, while
Bose-Einstein condensates can typically be created with up to
$10^7$ atoms every 10s, thus leading to a coherent matter wave
flux of $10^6$ atoms/s. The increase of coherent matter wave flux
is a major topic in nowadays coherent matter wave research, i.e.
by the development of novel atom laser sources discussed in
Sec.~\ref{s:atomlaser}.

Another important point are evolution time limitations due to the
influence of gravity on all kind of matter wave interferometer. In
many interferometric setups an increase in evolution time, i.e.
the time between splitting and recombining the interfering wave
packets, leads to a higher sensitivity. The precision of atom
interferometric measurements could be significantly increased
using a longer evolution time, which is mainly limited by two
factors: the spreading of the wave packet and the free fall
trajectory used in state-of-the-art instruments. One of the main
tasks is to exploit the superiority of coherent matter waves in
terms of their heisenberg limited momentum spread, which becomes
an advantage for evolution times of several seconds. In order to
obtain these times in the well understood free fall
interferometers, it seems promising to build a gravity free, i.e.
space bound, instrument. Another main route to enlarge the
evolution time consists in various efforts to develop matter wave
guiding structures as discussed in Sec.~\ref{s:guiding}. Both
routes are actively pursued in nowadays experimental activities.

Further promising attempts to reach increased precision with
matter wave interferometric sensors concentrate on obtaining
Heisenberg limited sensitivity, i.e. sensitivity going with
$\frac{1}{N}$ instead of the poissonian limit $\frac{1}{sqrt{N}}$,
with the number of interfering particles $N$. There are several
proposals to prepare squeezed and/or entangled matter wave states
which show the according
interference~\cite{Bouyer1997a,Pu2000a,Helmerson2001a,Sorensen2001a,Poulsen2001a,Duan2002a,Vogels2002a,Micheli2003a,Zhang2003a,Zhang2003b}.
Recent experiments demonstrated number squeezing of coherent
matter waves~\cite{Orzel2001a,Greiner2002a} as an important step
towards Heisenberg limited atom interferometry.

In the following we will discuss the implementation of atom
interferometers based on Bose-Einstein condensates as coherent
matter wave source. It is important to note, that all these
interferometers are based on single particle interference. Even if
a macroscopically ocupied Bose-Einstein condensate wavefunction is
used as a source, the beamsplitting process relies on single
particle processes. This is in analogy to optical setups using
laser sources, which are also based on single photon processes.

\subsection{Bragg atom interferometers with
BEC}\label{s:Bragginterferometer} The current development of atom
interferometers based on coherent matter waves aims at several
fundamental issues: using the low momentum spread, understanding
and controlling interparticle interactions and reaching Heisenberg
limited sensitivity.

One of the first interferometric schemes implemented with
Bose-Eistein condensates relies on Bragg diffraction for the
mirrors and
beamsplitters~\cite{Simsarian2000a,Torii2000a,Bongs2001a}. This
method (see Sec.~\ref{s:minimumuncertainty}) with a typical two
photon momentum transfer is especially interesting for matter
waves with a narrow momentum spread. The momentum spread of a
Bose-Einstein condensate wavefunction can be easily made smaller
than the Bragg diffraction momentum transfer. It is thus possible
to not only split the wavefunction in momentum but also in real
space, resulting in interferometers with well separated paths.
This allows to modify the phase along one of the paths with
suitable light pulses~\cite{Denschlag2000a}.

Bragg interferometers have been used in studies of the phase
evolution of BECs~\cite{Simsarian2000a} as well as their coherence
properties~\cite{Bongs2001a,Hellweg2003a}, in which they show
clear signatures of the nonlinear interparticle interactions of
the dense matter waves as well as phys fluctuations for finte
temperatures.

\begin{figure}
\begin{center}
\includegraphics[width=12cm]{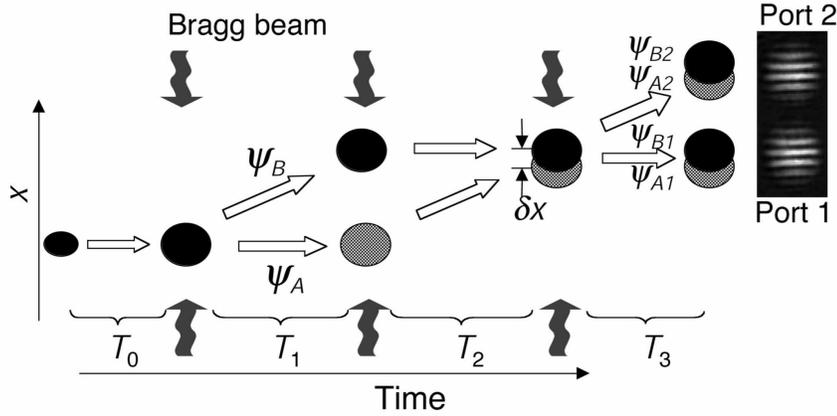}
\end{center}
     \caption{Bragg pulse interferometer based on a $\frac{\pi }{2}-\pi -\frac{\pi }{2}$ scheme.
     A free falling Bose-Einstein condensate wavefunction is subjected to three Bragg pulses. The first pulse is a
                 $\frac{\pi }{2}$ pulse splitting the condensate wavefunction into an equal superposition
                 of two different momentum states, leading to a subsequent spatial separation of the wavepackets.
                 The second pulse is a $\pi $ pulse inverting the states and thus acting as a mirror. The
                 third pulse is another $\frac{\pi }{2}$ beamsplitter pulse, recombining the wavepackets.
                 Interference leads to varying populations of the two different momentum state output
                 ports. These wavepackets in these ports are spatially separated after some time of
                 flight and can be individually detected. If the times between the Bragg pulses is equal, e.g. $t_1=t_2$
                 the interferometer is closed and varying
                 the interference phase leads to oscillation of the whole cloud between the output
                 ports. For the case of $t_1\ne t_2$, as in the example given here, the clouds
                 will only have partial overlap at the second beamsplitter making the observation of the phase
                 distribution of the condensate wavefunction
                 possible~\cite{Simsarian2000a}. Image with courtesy of National Institute of Standards and Technology, USA.}
     \label{fig:Bragginterferometer}
\end{figure}

A Bragg interferometer scheme is exemplarily shown in
Fig.~\ref{fig:Bragginterferometer}.

The above scheme with different evolution times between Bragg
pulses results in an open interferometer. In this case
Bose-Einstein condensates passing through this interferometer will
be only partially overlapped in the output ports. The resulting
interference signal allows a measurement of the condensate
autocorrelation function~\cite{Simsarian2000a}, which is modified
by interparticle interactions in the first stages of expansion
after release from the trap.

For the case that one is mainy interested in the spatial coherence
of the matter wave function, a simplified $\frac{\pi
}{2}$-$\frac{\pi }{2}$ pulse configuration can be
employed~\cite{Bongs2001a}. The first pulse splits the matter
wave, which is recombined by the second pulse after a variable
time, which defines the relative spatial shift of the interfering
wavepackets. This enables the measurement of the matter wave
coherence length.

The advantage of coherent matter waves in Bragg interferometric
schemes is not only due to the low momentum spread allowing
separated path interferometers, but is also reflected in the
clearly visible large scale density modulations in the output
ports. An advanced concept employing new readout schemes based on
this feature was recently demonstrated in a slightly different
interferometer type~\cite{Gupta2002a}. The scheme used in this
experiment employs a Kapitza Dirac pulse for initial splitting of
the matter wave, i.e. a very short standing wave light pulse
leading to three momentum states~\cite{Gould1986a}. A second order
Bragg pulse after some evolution time, $T$, redirects the split
wavepackets towards each other such that they pass through each
other at time $2T$. During this passage the coherent wavepackets
interfere and show the periodic buildup and decay of a matter wave
density grating. One intriguing feature of the interferometer
scheme is that the readout traces the reflection of an appropriate
laser beam off this density grating in time. This allows to
determine the interferometer phase in a single run, i.e. with one
wavepacket passing through. This interferometer demonstrates high
potential for precision measurements of $\frac{h}{m}$. It is
important to note that it only exploits the low momentum spread of
the Bose-Einstein condensate wavefunction but does not rely on its
coherence properties.

\subsection{Interferometers of Ramsey type}
Ramsey-type interferometers make use of the internal atomic
degrees of freedom in order to split manipulate and recombine a
wavepacket~\cite{Berman1997a}. The coupling can be realised with
or without momentum transfer, depending on the chosen interaction.
Ramsey-type interferometers have the advantage, that the different
paths of the wavepackets inside the interferometer as well as the
two output ports are represented by different internal states. As
a consequence neither a phase shift in one interferometer arm nor
the detection of the interferometry signal require a spatial
separation of the interfering wavepackets but only state selective
interactions. This feature makes Ramsey-type matter wave
interferometers very versatile and useful even for thermal
ensembles. They are implemented in most types of modern atom
interferometric precision sensors~\cite{Berman1997a}.

The high technological importance of this type of interferometer
gives rise to great interest concerning possible improvements with
coherent matter wave sources. A straightforward improvement lies
again in the low momentum spread, which allows a longer evolution
times with well defined trajectories. This essentially enhances
the flux along the desired pathways and minimises broadening
effects, which arise due to different phase contributions along
different paths or due to matter waves with different wavelength
contributing to the signal. In this sense these improvements are
analogous to the advantages of a laser source in optics.

The first implementation of a coherent matter wave Ramsey
interferometer was realised with a magnetically trapped
Bose-Einstein condensate at JILA~\cite{Hall1998b}. In this
experiment a $^{87}$Rb condensate was prepared in the
$|F=1,\,m_F=-1\rangle $ state and coupled to the
$|F=2,\,m_F=1\rangle $ state with two combined microwave and
radiofrequency fields. The interferometer scheme consisted in two
$\frac{\pi }{2}$ pulses (each transferring 50\% of the population
of one component into the other component and thus acting as 50\%
beamsplitter) separated by a variable evolution time. One aim of
this experiment was to study the effect of interparticle
interactions on the phase evolution of the interfering paths, in
particular looking for phase diffusion processes. It was found,
that the two matter wave components tend to spatially separate due
to their interactions, which in the trapping fields leads to
damped oscillatory motion relative to each other. The phase
evolution of the internal state superposition however remained
trackable without significant visible diffusion. This result shows
that the coherent evolution of a superposition of the internal
hyperfine ground states is not quickly destroyed due to
interparticle interaction and relative spatial movement of the two
components.

The interparticle interaction has however an important effect on
the frequency of the hyperfine transition. The density dependend
mean field energy difference for particles in one versus the other
spin state results in a position dependend shift of the resonance
frequency of the trapped ensemble.

The density dependence of the transition frequency difference was
employed in precise measurements of the difference of the
scattering lengths of the two involved spin
components~\cite{Harber2002a,Gorlitz2003a}. In similar experiments
with fermionic gases it was further shown, that these stay
noninteracting during the interferometric sequence. In this
respect incoherent fermionic matter waves are better suited for
high precision measurements, if the Fermi momentum spread of the
sample can be tolerated.

A further experiment with the $^{87}$Rb quasi-spin 1/2-system
investigated the influence of a position dependend
Rabi-frequency~\cite{Matthews1999a}. In this case a linearily
varying Rabi detuning was achieved by a slight spatial offset of
the centers of the $|F=1,\,m_F=-1\rangle $ component and the
$|F=2,\,m_F=1\rangle $ component. Coupling of the two components
for several Rabi oscillation cycles results in a complicated phase
pattern inscribed into the condensate wavefunction, which was
studied in detail in~\cite{Matthews1999a}.

The optimisation of interferometric schemes with respect to the
advantages of noninteracting fermionic atoms or low momentum
spread coherent bosonic matter waves is under current
investigation.

\subsection{Advanced concepts for atom optics with coherent matter
waves}\label{s:advanced}

The above matter wave interferometric schemes only make indirect
use of coherent matter waves by exploiting the possible low
momentum spread to reach high flux also for very long evolution
times. The resulting interferometry signal relies on single
particle interference effects. Its signal to noise ratio is
limited by counting statistics to a value proportional to
$\frac{1}{\sqrt{N}}$ with the number, $N$, of detected particles.
Advanced concepts aim at an increasing matter wave interferometer
sensitivity to the Heisenberg limit $\frac{1}{N}$. This might be
achieved by using squeezed matter waves similar to analogous
concepts in light optics or by arrangements employing entangled
particles or two coherent matter wave sources as discussed above.

\begin{figure}
\begin{center}
\includegraphics[width=8cm]{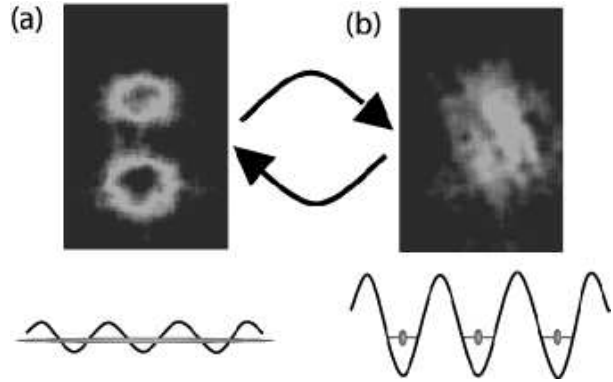}
\end{center}
     \caption{Transition from a coherent matter wave state (a) to a number
     squuzed state (b) with Bose-Einstein condensates loaded into an optical
     lattice. The absoption images show the interference/disappearance of interference,
     appearing when the matter wave is released from the lattice potential.
     The lower pictures show schematically how the matter wave distributes in the
     optical lattice, from delocalised to localised. Note that the transition is reversible.
     Experimental images with courtesy of M. Kasevich, Stanford university.}
     \label{f:squeezing}
\end{figure}

Experimentally important progress towards squeezed matter waves
has been made in optical lattice geometries, where number
squeezing has been observed~\cite{Orzel2001a,Greiner2002a}.
In~\cite{Orzel2001a} a Bose-Einstein condensate is loaded into a
1d standing wave optical dipole potential, coherently distributing
the condensate wavefunction over $\approx 30$ lattice sites. The
intensity of the light field is used to control the confinement at
each site and the tunneling rate between adjacent sites. For very
high intensities the limit of negligible tunneling versus the on
site interparticle interaction is reached. This leads to a quantum
phase transition to an insulating state similar to the Mott
insulator discussed in Sec.~\ref{s:Mottinsulator}. The special
feature of this state is that the number of atoms in each
individual potential well is fixed to a precise value, i.e. the
atom number is squeezed. As the atom number is intrinsically
related to the phase uncertainty by a phase-number uncertainty
relation, the phase shows maximum uncertainty in this state. The
transition to the number squeezed state with high phase
uncertainty is clearly visible in the loss of interference
contrast of the matter wavepackets, when released from the optical
lattice (see Fig.~\ref{f:squeezing}).

One idea is to enhance the sensitivity of interferometric
measurements by transforming this number squeezed state into a
phase squeezed state. This phase squeezed state can then be
employed in a contrast interferometer scheme to reach Heisenberg
limited sensitivity~\cite{Bouyer1997a}. First experiments using
fast intensity changes of the optical lattice state oscillations
from number squeezing to phase squeezing, which open interesting
perspectives for the realisation of such a scheme.

Furthermore it should be noted, that the progress in manipulating
the external and internal degrees of freedom of matter waves
enables the creation of entangled atomic pairs due to collisional
processes~\cite{Helmerson2001a,Vogels2002a,Mandel2003a}. Feeding a
matter wave interferometer with these entangled pairs represents
another way of reaching Heisenberg limited sensitivity (see
e.g.~\cite{Dunningham2002a}.

Modern atom interferometric schemes allow highest precision in
several types of measurements but typically do not directly
benifit from the coherence of a matter wave source. The optimum
configuration and possible improvements with novel coherent
sources is subject of ongoing research. Furthermore the high
experimentally achievable control over matter wave internal and
external dynamics opens new possibilities for reaching ultimate
sensitivity with adapted sources.

\section{Coherent matter waves in optical lattices}
The combination of coherent matter waves with periodic potential
structures opens a new quality in controlled matter wave
manipulation and in the realisation of model systems for condensed
matter phenomena. Optical lattices created from interfering laser
beams are defect-free periodic potentials with easily tunable
parameters. Depending on the number and the geometry of
interfering lasers 1d, 2d or 3d lattices can be created with
different symmetries~\cite{Petsas1994a}. Changing the laser
intensity the system can be tuned from the tight binding model to
a free sample, relative laser frequency detuning allows to exert
precise accelerations on the sample and different frequency or
different polarisation lattices can shift the relative position of
different components in multi-component matter waves in a
controlled way.

\subsection{Coherent phenomena}
The adiabatic transfer of a Bose-Einstein condensate into a
periodic potential as created by an optical lattice gives rise to
a macroscopic population of the lowest momentum state in the
lowest Bloch band~\cite{Greiner2001a,Greiner2001b}. If tunneling
between lattice sites is not negligible, i.e. the Mott insulator
regime is not realised, the coherence of the matter wave will be
preserved. This coherence leads to various interference effects,
e.g. in transport phenomena and in free expansion after switching
off the lattice potential. The possibility to project the momentum
distribution of the matter wave inside the lattice onto free space
momentum states by abruptly switching off the optical potentials
offers unique possibilities to study coherent effects in crystal
lattices. This type of investigation of condensed matter phenomena
opens new pathways, which are not accessible in real condensed
matter systems.
\begin{figure}
\begin{center}
\includegraphics[width=12cm]{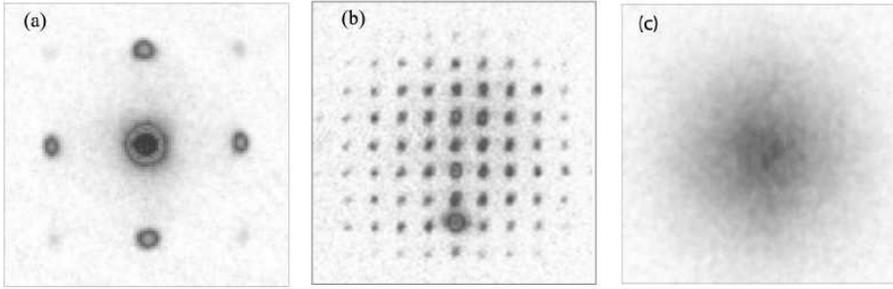}
\end{center}
     \caption{Interference of coherent matter waves interacting with a three dimensional
     optical lattice. The absorption images were taken after a time of flight period, such that
     the visible distribution corresponds to the momentum distribution of the ensemble.
     (a) Interference peaks from a matter wave adiabatically loaded into
     the lattice, which was then abruptly switched off. In this case only a few momentum
     components are occupied. (b) Interferences for a matter wave subjected to a pulsed
     optical lattice potential with many occupied momentum components. (c) Same as (a) but with
     a high optical lattice confinement, leading to a Mott insulator state
     (see Sec.~\ref{s:Mottinsulator}). In this state the inter-site coherence is completely lost
     and the image shows no interferences.  Note that the
     momentum separation of the peaks in (a) and (b) is the same, only the scales are
     different.(c) uses the same scale as (a).
     Images with courtesy of I. Bloch, Mainz university and T. W. H\"ansch, MPQ.}
     \label{f:latticeinterference}
\end{figure}
The interference structure obtained some time after switching off
a 3d optical lattice (see Fig.~\ref{f:latticeinterference}) can be
interpreted in several ways: First one can use a condensed matter
approach in considering the pattern as the momentum distribution
of the lowest state in the first Brillouin zone
(Fig.~\ref{f:latticeinterference} (a)). The crystal momentum
distribution peaked at zero momentum is periodically repeated in
free space momentum space with a spacing proportional to the
inverse lattice period.

A second picture of the process comes from the analogy to wave
diffraction by periodic structures, e.g. Laue peaks from electron
diffraction by crystal structures. The idea is to associate the
matter wave inside the optical lattice with waves passing through
a crystal and later interfere. This picture clearly incorporates
interference effects but it should be noted that it rather
corresponds to a pulsed matter waver - optical lattice interaction
(Fig.~\ref{f:latticeinterference} (b)).

\begin{figure}
\begin{center}
\includegraphics[width=12cm]{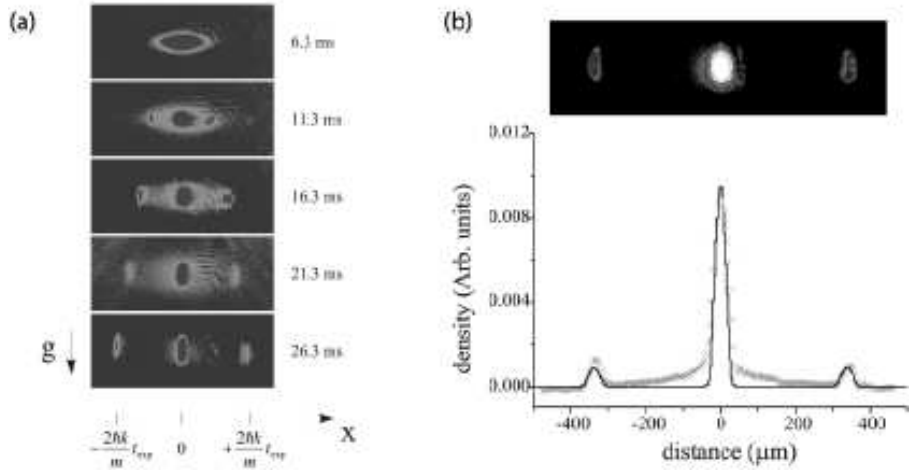}
\end{center}
     \caption{Interference structures of a coherent matter wave released
     from a one dimensional optical lattice (which is abruptly switched off).
     (a) Time of light development of the ensemble with the buildup of interferences
     between the different momentum components. (b) Comparison of the interference
     profile with the theoretical prediction for an array of expanding condensates~\cite{Pedri2001a}.
     Images with courtesy of M. Inguscio, LENS, Florence.}
     \label{f:latticeinterference1d}
\end{figure}

The above ideas also hold for the case of a 1d optical lattice,
where an analogy to laser beam diffraction off a periodic grating
can be made. Recent experiments have analysed the according matter
wave interference structure and found similar structures to the
ones obtained in optical grating diffraction
experiments~\cite{Pedri2001a} (see
Fig.~\ref{f:latticeinterference1d}). These structures can be
modified by moving or accelerating the lattice. Accelerating a
lattice to a certain velocity corresponds to imposing a linear
phase gradient across the matter wave. In the optics picture this
mimics a laser beam incident on a diffraction grating with a
non-orthogonal angle between grating and laser propagation
direction. This regime has first been explored by the group of M.
Kasevich at Yale university with a 1d lattice oriented in the
direction of gravity and might be useful for high precision
measurements of accelerations.

\subsection{Bloch oscillations and Josephson junctions}
Bloch oscillations are a paradigm of condensed matter physics and
intrinsic to the Bloch lattice model. The acceleration of
particles in the lowest momentum states of a perfect lattice leads
to an increase in crystal momentum until the border of the first
Brillouin zone is reached. A further acceleration leads to the
transfer of the particles to the opposite border of the Brillouin
zone with the inverse momentum. From there the particles start
again gaining crystal momentum, until they reach the border of the
Brillouin zone another time and everything starts over. Applying a
constant force to particles confined in an ideal lattice thus
leads to a sawtooth-like periodic increase and inversion of
momentum.

In a matter wave picture the acceleration corresponds to an
increase of the wavenumber of the particles. The border of the
Brillouin zone is reached, when this wavenumber equals the $\pi $
over the lattice period. In other terms the Bragg scattering
condition for the matter waves to be reflected by the lattice is
fulfilled and the velocity of the matter wave is inverted.

Due to lattice defects and phonon excitations Bloch oscillations
are very difficult to observe in condensed matter systems. The
implementation of an atom optical model system with atomic
particles in a practically defect-free optical lattice has changed
this situation and enabled the observation of matter wave Bloch
oscillations~\cite{Bendahan1996a}. The first experiments were
based on laser-cooled thermal ensembles and no spatial information
on the Bloch oscillations could be obtained. In recent experiments
based on coherent matter waves both the momentum as well as the
spatial evolution of the wavepackets were
investigated~\cite{Morsch2001a}. With these experiments the
intrinsic features of Bloch oscillations became clearly visible in
time of flight images.

While Bloch oscillations typically occur in relatively shallow
lattices with significantly broadened conduction bands the related
physics of Josephson junctions is based on tunneling through a
relatively heigh barrier as might be realised in a deep lattice.
Both regimes give rise to oscillatory behaviour and a smooth
crossover can be expected in the lattice realisation.

In experiment a vertical array of Josephson junctions was realised
by loading a vertical 1d optical lattice with a Bose-Einstein
condensate~\cite{Anderson1998a}. This setup incorporates output
coupling of the matter wave due to tunneling through the potential
barriers and thus realises an atom laser. Due to the gravitational
potential difference the phase between different sites increases
linearly, which leads to an oscillating Josephson current. The
atom laser output is pulsed according to this oscillation. The
atom laser repetition frequency is a means to measure gravity with
prospects for high precision measurements.

Another fascinating feature of Josephson tunneling was
demonstrated by comparing the movement of the normal (thermal)
component and the wavefunction of a Bose-Einstein
condensate~\cite{Cataliotti2001a}. A Bose-Einstein condensate with
significant normal component was loaded into a 1d optical lattice
with a superposed harmonic oscillator potential. The lattice
potential was relatively high compared to the temperture of the
sample but still allowed significant tunneling. The motion of the
ensemble was investigated after a sudden displacement of the
harmonic oscillator potential minimum. The ''classical'' normal
component was essentially confined to its original position, while
the superfluid condensate wavefunction fulfilled oscillations due
to coherent Josephson tunneling through the potential barriers. An
analysis of the potential dependend oscillation period is in
agreement with the prediction from Josephson junction
theory~\cite{Josephson1962a}. This experiment shows the
extraordinary properties of superfluidity in an impressive way and
is analogous to experiments on the flow of superfluid helium in
porous vycor.
\begin{figure}
\begin{center}
\includegraphics[width=12cm]{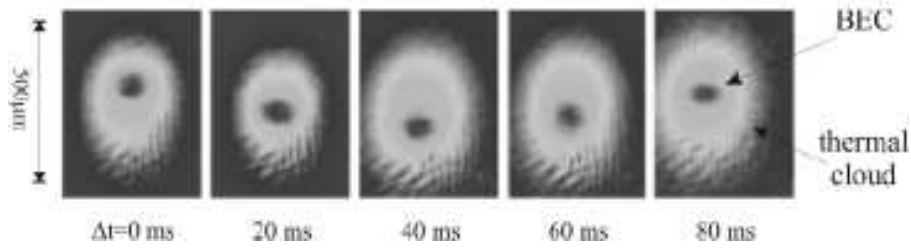}
\end{center}
     \caption{Motion of a Bose-Einstein condensate with significant normal
     component (''thermal cloud'') in a one dimensional optical lattice with
     superposed harmonic oscillator potential. In the beginning the
     Bose-Einstein condensate was displaced from the minimum of the harmonic potential.
     Due to the optical lattice only coherent tunneling currents are possible, resulting
     in an oscillatory motion of the condensate wavefunction, while the normal component
     remains fixed in space. Image with courtesy of M. Inguscio, LENS, Florence.}.
     \label{f:josephson1}
\end{figure}

\subsection{Mott insulator transition}\label{s:Mottinsulator}
The recent observation of the superfluid to Mott insulator quantum
phase transition demonstrates the power of coherent matter waves
in optical lattices as model system for condensed matter phenomena
in an impressive way~\cite{Greiner2002a}. A quantum phase
transition is special in the sense, that it is driven by quantum
instead of classical fluctuations. Furthermore it is based on a
competition between two different contributions in the hamiltonian
of the system instead of a competition between inner energy and
entropy~\cite{Sachdev2001a}.

The superfluid to Mott insulator transition of coherent matter
waves in an optical lattice is driven by the competition between
on-site interparticle interactions and the tunnel coupling of
adjacent sites. The system consists of bosonic atoms in a
relatively deep 3d optical lattice, such that site-to-site motion
is solely based on tunneling processes. The according second
quantised hamiltonian in the Bose-Hubbard
model~\cite{Fischer1989a} without external confinement is given
by~\cite{Greiner2002a}
\begin{equation}
  H=-J\sum_{\langle i,j \rangle }\hat{a}_i^{\dagger }\hat{a}_j +
  + \frac{1}{2}U\sum_i \hat{n}_i\left(
  \hat{n}_i-1 \right) .
\end{equation}
The number operator $\hat{n}_i = \hat{a}_i^{\dagger }\hat{a}_i $
is defined by the bosonic operators $\hat{a}_i^{\dagger }$ and
$\hat{a}_i$ creating or annihilating a particle at lattice site
$i$. The energy scales of the tunneling and interaction are given
by the hopping matrix element $J$ and the on-site interaction
matrix element $U$ (see~\cite{Greiner2002a}).

In the limit of high tunneling as compared to the on-site
interaction, $\frac{J}{U}\gg 1$, long range coherence can be
maintained. The many-body ground state wavefunction is then given
by a multiple occupied single particle wavefunction similar to the
case of a harmonically trapped Bose-Einstein condensate. Due to
the long range coherence this state corresponds to a superfluid
inside the optical lattice (see Fig.~\ref{f:latticeinterference}
(a)).

In the opposite limit, $\frac{J}{U}\ll 1$, in which tunneling is
negligible long range coherence is lost and the particles become
localised at single lattice sites. For commensurate fillings the
many body ground state is represented by a product wavefunction of
single site Fock state wavefunctions with exactly the same
particle number at each site. In this sense this regime again
realises particle number squeezing as discussed in
section~\ref{s:advanced}. In this interpretation the total phase
uncertainty due to phase fluctuations in between different sites
suppresses transport and the system becomes an insulator (see
Fig.~\ref{f:latticeinterference})(c)).

The transition from superfluid tunneling transport to an insulator
regime can also intuitively be explained in a picture based on the
energy levels of the individual wells. Considering one atom per
lattice site, the lowest energy level of each site will be singly
occupied. Furthermore, regarding the entire optical crystal
system, there will be an energy band formed by the combination of
all these local energy levels with a broadenening according to the
inter-site tunneling. Transport of atoms from one lattice site to
another would in a first step lead to a double occupancy of one
site. In this case the local energy level would be shifted by the
inter-particle interaction energy, $U$. Following energetic
considerations transport can only occur, if the tunnel
broadenening is larger than the interparticle interaction term,
i.e. $J > U$. The Mott insulator state is reached when the
tunneling energy cannot account for the interaction energy
difference any more, $J<U$.

In addition this simplified intuitive picture directly leads to
the occurrence of an energy gap for transport excitation of the
sample, once the insulating state is reached. The experimental
investigation of the above superfluid to Mott insulator transition
\cite{Greiner2002a} confirmed this energy gap and its dependence
on the relative values of tunneling matrix element and interaction
energy.

The very clean realisation of the Mott insulator starting with
coherent matter waves in an optical lattice was a milestone in
modelling condensed matter phenomena. In this sense the well
controlled systems for coherent matter wave creation and control
represent an analog quantum computer which directly maps one
hamiltonian onto another.

In addition the ultimate control over single atoms reached in the
above experiment was further extended to different internal states
with controllable interactions~\cite{Greiner2002b,Mandel2003b} new
interferometer schemes and multi-particle entanglement at
will~\cite{Mandel2003a}. This paves the way for new types of
matter wave sensors, quantum computers and controlled quantum
chemistry.

\section{Concluding remarks}
This article is intended to give a representative overview of the
field of coherent matter waves with emphasis on the physics with
this new type of system. Due to the rapid developments and the
explosion in activities and publications following the first
realisation of Bose-Einstein condensation in dilute atomic gases a
comprehensive coverage of the entire field is beyond the scope of
a single review article. For further information we refer the
reader to the available
books~\cite{Griffin1995c,Martellucci2000a,Savage2001a,Pethick2002a,Pitaevskii2003a}
and review articles with different
emphasis~\cite{Cornell1996a,Burnett1999a,Cornell1999a,Ketterle1999a,Dalfovo1999a,Griffin1999b,Helmerson1999b,Courteille2001a,Stamper-Kurn2001a,Ketterle2001a,Cornell2002a,Cornell2002b,Bongs2003a}.

We have shown that the ability to produce macroscopically occupied
matter wave functions via Bose-Einstein condensation is the basis
for many new insights into the physics of coherent matter waves.
For the first time it is now possible to directly image the
probability distribution of a matter wavefunction onto a CCD
camera. Although this distribution is integrated along the
detection direction one obtains spatially resolved two dimensional
information, which greatly influenced our understanding of quantum
physics.

As an example discussed in Sec.~\ref{s:interferece}, this spatial
resolution enabled the observation of the interference between two
Bose-Einstein condensate wavefunctions in a single shot. The
intense theoretical discussion on this interference pattern
established new views on how Fock state wavefunctions with total
phase uncertainty can nevertheless interfere, e.g. by the
establishment of a relative phase during the detection process.

In addition the spatial resolution can also be used for local
manipulations of the quantum mechanical wave function, e.g. by
phase imprinting discussed in Sec.~\ref{s:phasemanip}. This novel
access to quantum physics demonstrated its power in the creation
of dark matter wave solitons as basic nonlinear excitation in
these systems. Further examples of the local manipulation of the
condensate wavefunction are the demonstration of its superfluid
character and the creation of quantised vortices.

These examples are just the starting point for the field of
quantum engineering with coherent matter waves which promises many
more insights into the quantum world.

As an important point for future developments we discussed the
achievements and new schemes in the development of new sources for
coherent matter waves. One major goal in this respect is the
realisation of a continous atom laser with high flux. Next to
experimental activities to reach this goal there are ongoing
studies on the coherence properties of such a device. First
experimental studies on coherence of an atom laser source were
already possible based on pulsed or quasi-continous atom lasers.
These were realised by controlled output coupling of Bose-Einstein
condensates confined in a trapping potential.

We presented several intriguing aspects connected to the
nonlinearity intrinsic to matter waves due to the interparticle
interactions. These aspects highlight the analogy to optical laser
fields, with phenomena such as four wave mixing and solitons. The
interactions connecting the internal atomic spin degrees of
freedom however also allow novel effects, such as magnetism and
coherent spin oscillations, in these fascinating quantum gases.
Further novel developments aim at two-particle entanglement
created during the nonlinear matter wave interactions.

As a further point we discussed the status of development of
coherence preserving atom optical elements, necessary to
manipulate coherent matter waves. In particular we concentrated on
optical and magnetic approaches to guiding structures as well as
different types of mirrors and beamsplitters.

We note in this article that many promising applications for
coherent matter waves, such as high precision measurements,
primarily require a high flux. For state-of-the-art
interferometric setups the main benefit from the coherence
properties is the brightness of the source. As discussed in
Sec.~\ref{s:atominterferometry} there are however limitations due
to the gravitational acceleration, which in practise limit the
evolution time, such that the gain versus a thermal source is
relatively small. Furthermore the one has to control intrinsic
nonlinearity of nowadays relatively dense coherent matter wave
sources in order to achieve high precision interferometric
measurements.

We discussed several types of atom interferometric setups realised
with Bose-Einstein condensate wavefunctions, in particular Bragg
and Raman interferometer schemes. These setups presented important
steps towards the understanding of the behaviour of coherent
matter waves in interferometric applications. In addition they
added significant information on the coherence properties and
phase of the condensate wavefunction. We also discussed
experiments and ideas on advanced approaches to coherent matter
wave interferometry, which make use of guiding structures,
squeezing or multi-particle entanglement.

As final points we concentrated on recent achievements with
coherent matter waves in optical lattices. These systems are
particularly interesting in terms of model systems for condensed
matter phenomena. Phenomena such as Bloch oscillations, Josephson
currents and Mott insulator phases have been realised in a very
clean way. They mark the beginning of this novel direction in
physics with coherent matter waves.

Another important aspect of matter wave systems inside optical
lattice structures lies in the futher increased control one can
exert over the system. We discussed progress towards moving single
atoms at will in momentum and position space, letting them
interact in a well defined way and detecting their individual
state.

In summary the physics of coherent matter waves rapidly grows into
diverse fields. Starting from fundamental insights into the
quantum world we find fascinating analogies to coherent and
nonlinear optics, promising applications in high precision
measurements, intriguing possibilities to model condensed matter
phenomena and novel prospects for quantum computation. Although
several basic features have been investigated the whole field is
still at the beginning of many interesting developments and
clearly continues to expand. One can expect exciting results and
fundamental new physics in near future in the area of the physics
with coherent matter waves.

\bibliographystyle{unsrt}
\bibliography{ROP}

\end{document}